%% file: main.tex
\documentclass[sigconf]{acmart}
\setcopyright{none}
\renewcommand\footnotetextcopyrightpermission[1]{}

\AtBeginDocument{%
  \providecommand\BibTeX{{%
    Bib\TeX}}}

\def\BibTeX{{\rm B\kern-.05em{\sc i\kern-.025em b}\kern-.08emT\kern-.1667em\lower.7ex\hbox{E}\kern-.125emX}}

\usepackage{makecell}
\usepackage{booktabs}
\usepackage{graphicx}
\usepackage{bookmark}
\usepackage[normalem]{ulem}
\usepackage{tcolorbox}
\usepackage{pifont}
\usepackage{xcolor}
\definecolor{myblue}{RGB}{0,90,150}
\usepackage{nicefrac}
\usepackage{siunitx}
\usepackage{array,framed}
\usepackage{cleveref}
\crefformat{section}{\S#2#1#3} 
\crefformat{subsection}{\S#2#1#3}
\crefformat{subsubsection}{\S#2#1#3}
\usepackage{
  color,
  float,
  epsfig,
  wrapfig,
  graphics,
  graphicx,
  subcaption
}
\usepackage{textcomp}
\usepackage{setspace}
\usepackage{latexsym,fancyhdr,url}
\usepackage{enumerate}
\usepackage{algorithm2e}
\usepackage{algpseudocode}
\usepackage{graphicx} 
\usepackage{xparse} 
\usepackage{xspace}
\usepackage{multirow}
\usepackage{csvsimple}
\usepackage{balance}

\tcbset{
  simplechat/.style={
    colback=gray!7!white,
    colframe=black,
    boxrule=0.4pt,
    arc=1.5mm,
    boxsep=0pt,
    left=3pt,
    right=3pt,
    top=3pt,
    bottom=3pt,
    width=0.95\columnwidth
  }
}

\usepackage{
  tikz,
  pgfplots,
  pgfplotstable
}
\usepackage{hyperref}

\usetikzlibrary{
  shapes.geometric,
  arrows,
  external,
  pgfplots.groupplots,
  matrix
}

\pgfplotsset{compat=1.9}

\usepackage{mathtools}

\DeclareMathAlphabet{\mathcal}{OMS}{cmsy}{m}{n}

\usepackage{epstopdf}

\DeclareGraphicsExtensions{%
    .png,.PNG,%
    .pdf,.PDF,%
    .jpg,.mps,.jpeg,.jbig2,.jb2,.JPG,.JPEG,.JBIG2,.JB2}
    
\input{Main_document/defs}
\newif\ifDEBUG
\DEBUGtrue         

\ifDEBUG

  \usepackage{xcolor} 
  \newcommand{\Priya}[1]{\textcolor{magenta}{\textbf{[Priya: #1]}}}
  \newcommand{\SR}[1]{\textcolor{blue}{\textbf{[Sazzadur: #1]}}}
  \newcommand{\Sonja}[1]{\textcolor{green!50!black}{\textbf{[Sonja: #1]}}}
  
  \newcommand{\JC}[1]{\textcolor{orange}{\textbf{[Jesse: #1]}}}
  \newcommand{\RY}[1]{\textcolor{cyan!60!black}{\textbf{[Rubin: #1]}}}
\else
  \newcommand{\Priya}[1]{}
  \newcommand{\SR}[1]{}
  \newcommand{\Sonja}[1]{}
  \newcommand{\rakib}[1]{}
  \newcommand{\JC}[1]{}
  \newcommand{\RY}[1]{}
\fi

\begin{document}
\fancyhead{}
\title{An Evaluation of Chat Safety Moderations in Roblox} 
\def\thetitle{An Evaluation of Chat Safety Moderations in Roblox}
\title{\thetitle}




\author{Priya Kaushik}
\affiliation{%
  \department{\textit{Department of Computer Science}}
  \institution{\textit{University of Arizona}}
  \city{Tucson, Arizona}
  \country{USA}
}
\email{priyakaushik@arizona.edu}
\author{Sonja Brown}
\affiliation{%
  \department{\textit{Department of Computer Science}}
  \institution{\textit{University of Arizona}}
  \city{Tucson, Arizona}
  \country{USA}
}
\email{sonjabrown@arizona.edu}
\author{Rakibul Hasan}
\affiliation{%
  \department{\textit{Department of Computer Science}}
  \institution{\textit{Arizona State University}}
  \city{Tempe, Arizona}
  \country{USA}
}
\email{rhasan3@asu.edu}

\author{Sazzadur Rahaman}
\affiliation{%
  \department{\textit{Department of Computer Science}}
  \institution{\textit{University of Arizona}}
  \city{Tucson, Arizona}
  \country{USA}
}
\email{sazz@arizona.edu}

\date{}

\input{Main_document/abstract}

\maketitle


\input{Main_document/intro}

\input{Main_document/data_meth}
\input{Main_document/discussion}

\input{Main_document/relatedwork}

\input{Main_document/conclusion}


\bibliographystyle{ACM-Reference-Format}
\bibliography{bibliography}

\appendix
\input{Main_document/appendix}
\end{document}

%% file: Main_document/defs.tex
\usepackage{xparse}
\newcommand{\bnm}{\begin{newmath}}
\newcommand{\enm}{\end{newmath}}

\newcommand{\bea}{\begin{eqnarray*}}%
\newcommand{\eea}{\end{eqnarray*}}%

\newcommand{\bne}{\begin{newequation}}
\newcommand{\ene}{\end{newequation}}

\newcommand{\bal}{\begin{newalign}}
\newcommand{\eal}{\end{newalign}}

\newenvironment{newalign}{\begin{align}%
\setlength{\abovedisplayskip}{4pt}%
\setlength{\belowdisplayskip}{4pt}%
\setlength{\abovedisplayshortskip}{6pt}%
\setlength{\belowdisplayshortskip}{6pt} }{\end{align}}

\newenvironment{newmath}{\begin{displaymath}%
\setlength{\abovedisplayskip}{4pt}%
\setlength{\belowdisplayskip}{4pt}%
\setlength{\abovedisplayshortskip}{6pt}%
\setlength{\belowdisplayshortskip}{6pt} }{\end{displaymath}}

\newenvironment{newequation}{\begin{equation}%
\setlength{\abovedisplayskip}{4pt}%
\setlength{\belowdisplayskip}{4pt}%
\setlength{\abovedisplayshortskip}{6pt}%
\setlength{\belowdisplayshortskip}{6pt} }{\end{equation}}

\newcounter{ctr}

\newcounter{mytable}
\def\mytable{\begin{centering}\refstepcounter{mytable}}
\def\endmytable{\end{centering}}

\newcounter{myfig}
\def\myfig{\begin{centering}\refstepcounter{myfig}}
\def\endmyfig{\end{centering}}

\newlength{\saveparindent}
\newlength{\saveparskip}
\newcommand{\E}{{\rm I\kern-.3em E}}

\renewcommand{\eqref}[1]{\mbox{Equation~(\ref{#1})}}

\def \part {part}

\renewcommand{\paragraph}[1]{\vspace*{6pt}\noindent\textbf{#1}\;}

\def \blackslug{\hbox{\hskip 1pt \vrule width 4pt height 8pt
    depth 1.5pt \hskip 1pt}}
\def \qed{\quad\blackslug\lower 8.5pt\null\par}

\newcounter{mynote}[section]

\newcommand\ignore[1]{}

\newcounter{rcnote}[section]

\newcounter{mrnote}[section]

\newcounter{fknote}[section]

\newcounter{anote}[section]

\DeclareMathSymbol{\mlq}{\mathord}{operators}{``}
\DeclareMathSymbol{\mrq}{\mathord}{operators}{`'}

\newcommand{\rhf}[2]{R_{f, \gamma}}

\DeclareDocumentCommand{\edist}{o o}{
  \ensuremath{
    \IfNoValueTF{#1}{{d}}{{\sf d}(#1,#2)}
  }
}

\newcommand{\olrk}[1]{\ifx\nursymbol#1\else\!\!\mskip4.5mu plus 0.5mu\left(\mskip0.5mu plus0.5mu #1\mskip1.5mu plus0.5mu \right)\fi}

\NewDocumentCommand{\indseq}{ O{1} O{r} }{{#1}\ldots {#2}}


%% file: Main_document/abstract.tex
\begin{abstract}

Roblox is among the most popular online gaming platforms, used by hundreds of millions of users every day. A substantial portion of these users are underage, who are at a greater risk, where abusive users may utilize Roblox's real-time chat interface to make the initial contact with potential victims. Roblox employs automated chat moderation mechanisms to detect potentially abusive messages; however, to date, their effectiveness has not been independently investigated. Toward this goal, we collected \textbf{approximately 2 million} chat messages from four games across multiple age groups and analyzed them to evaluate the moderation system. These messages were collected from public game servers following ethical and legal norms as well as Roblox's terms of service.

We use this corpus to qualitatively study which types of unsafe chats escape the moderation system and how policy-violating users evade the moderation system. Given the dataset's scale, it is prohibitively expensive to conduct qualitative content analysis manually. Therefore, we adopt a two-step approach. First, we manually labeled safe and unsafe messages (n=99.8K) and used them as a ground truth to evaluate four locally hosted state-of-the-art large language models (LLMs). Next, the best-performing LLM was applied to the entire corpus to identify potentially unsafe messages, which we manually categorized using iterative open and axial coding methods until thematic saturation was reached. Overall, our findings reveal a troublesome reality: numerous instances of unsafe chat messages related to grooming, sexualizing minors, bullying, \& harassment, violence, self-harm, and sharing sensitive information, etc., escaped the current moderation. Our analysis of users whose messages were previously flagged revealed that they continue to send harmful messages by employing a wide range of techniques to evade the moderation system.

\textcolor{red}{[WARNING: This paper includes examples of self-harm, violence, sexually explicit, and profane content that may be triggering.]}

\end{abstract}

%% file: Main_document/intro.tex
\section{Introduction}
\label{sec:intro}

Massively multiplayer online (MMO) games are large social ecosystems that allow users to interact and build digital adventures with others around the world. These platforms mainly attract underage players (ages 17 and below) and transform online video games into a venue for youth socialization~\cite{Statista_2025}. Roblox is among the largest child-oriented MMO platforms with nearly 144M daily active users worldwide~\cite{clement2025roblox}. Unfortunately, these platforms can also become a breeding ground for harmful interactions for minors, such as grooming and exploitation. Incidents like the California Kidnapping case~\cite{Tan_2025} highlight that interactions initiated on platforms like Roblox can escalate beyond the virtual environment and may lead to real-world harms. Roblox emphasizes strict child safety standards~\cite{robloxsafety}, and includes a real-time chat moderation system that scales at billions of messages per day~\cite{Koneru_2025} to detect and filter inappropriate content~\cite{robloxRobloxCommunity}. 


Unfortunately, even with such protective mechanisms, concerns about safety, privacy, and exploitation continue to rise. Recent lawsuits filed in the United States and reports in the media raise concerns about the effectiveness of Roblox's moderation system, especially in relation to grooming behavior and off-platform pivoting~\cite{LACountyRoblox2026, seitz, Carville_DAnastasio_2024, Ellery_2025}. Unlike explicit profanity or hate speech, grooming evolves slowly and varies in style, duration, and intensity~\cite{OSG_Parcelli}. Once trust is established, individuals move to extract personalized information, migrate to external platforms, or even arrange to meet in person~\cite{Chawki_2025}. 

Against this backdrop, it is both urgent and critical to scientifically evaluate Roblox's chat safety moderation~\cite{Wrocherinsky_2023}. In particular, we lack understanding of what types of policy violations moderation systems capture or miss, and how conversations evolve to extract personal information or pivot off-platform. To address this gap, we conduct a large-scale measurement study of over two million chat messages across four live interactive Roblox games for two different age groups (\textit{9+} and \textit{13+}). In this study, we answer the following research questions:
\begin{itemize}\label{RQs}
    \item \textbf{RQ1}: \label{RQ1}What child-safety violations are bypassing Roblox's chat moderation system?
    \item \textbf{RQ2}: \label{RQ2}What strategies are policy-violating users employing to intentionally bypass the chat moderation system?
\end{itemize} 

Answering these questions required us to overcome a number of logistical and technical challenges. Unlike other previously studied platforms (e.g.,~\cite{gatta2023interconnected, goldstein2023understanding}), Roblox does not provide any API for monitoring or collecting user content. Moreover, chat messages on Roblox are rendered as part of the game environment and disappear after some time, which makes it crowdsourcing data from users infeasible~\cite{gonccalves2026potential, razi2023sliding}. Furthermore, collecting and analyzing such datasets have ethical implications. Thus, we collaborated with our institutional ethics board to develop a protocol for data collection and analysis, which we summarize below (see \cref{ethics} for details). 

For ethical concerns, we avoided reverse-engineering proprietary game binaries, and relied on video recordings of live gameplay from public servers in a \textit{non-intrusive} manner. We then built an OCR (Optical character recognition)-based framework to extract messages from these videos. As game elements interfere with the chat messages, they add noise and lower the OCR performance. To prevent this, we draw inspiration from metamorphic testing~\cite{DBLP:journals/corr/abs-2002-12543}, in which variations of the original image are created through careful transformations to assess the OCR model's output consistency and confidence. To ensure a baseline quality, we exclude inconsistent and low-confidence outputs in our final corpus. \textit{It is worth noting that our framework is platform-agnostic and can be easily used to collect a real-time chat message corpus from any platform in a non-intrusive manner.} In the end, we collected \textbf{approximately 2 million (2,040,839)} chat messages from \textbf{105,214} users across 4 popular Roblox games.

To understand which harmful messages escape chat moderation (\textbf{RQ1}), we need to identify messages that violate Roblox's chat safety guidelines~\cite{robloxcomm}. 
Manual labeling of such a large dataset is infeasible; thus, we adopted a semi-automated approach. First, we labeled 2,000 \textit{conversations} (blocks of 50 messages) and use them as ground truths to test several state-of-the-art large language models using a few-shot prompting strategy. All models were locally hosted to avoid uploading data to third-party cloud servers. After evaluation, the best performing model, (\textit{gpt-oss-120b}~\cite{agarwal2025gpt}), was applied to the full corpus to identify potentially harmful content. The model identified 7,254 conversations to be unsafe. 
We classified these messages into multiple categories based on the LLM's reasoning outputs. Next, we use a saturation-based qualitative thematic analysis to find out (1) if a given message is truly unsafe; (2) identify the specific factors that make it problematic. Note that while this model-assisted pipeline reduces annotation efforts, it may miss subtle or rare unsafe categories due to (1) false negatives in the initial model-based filtering and (2) saturation-driven analysis, where saturation is reached once no new categories emerge from the sampled data, potentially overlooking low-frequency phenomena outside the observed sample. We argue that this pipeline is reasonable for capturing the dominant patterns of unsafe behavior at scale, even if it does not exhaustively enumerate all edge-case categories. 

Our analysis revealed a troublesome reality. We found Roblox's moderation system, while filtering isolated explicit terms, grossly fails to detect unsafe content that emerges over the course of conversations. Even for the filtered one, often the intent was leaked. Thus, generally unsafe content remained visible in most of the categories, such as grooming, sexually explicit content, harassment, PII leakage, and violent language, that Roblox's moderation system aims to filter. For a child-centered platform, this is particularly concerning, as benign social interactions can evolve into harmful trajectories involving personal information disclosure, coercion, and off-platform migration -- opening potentials for wide ranges of physical, sexual, and mental abuses.

Next, we examined what strategies users adopted to evade moderation (\textbf{RQ2}). Specifically, we focus on users who have been moderated before (i.e., their messages have been filtered by Roblox's moderation system) to examine if and how they subsequently tried to evade moderation.  
This analysis was guided by the hypothesis that \textit{users who exhibited violations are more likely to employ or evolve evasion strategies, making them a representative and information-rich subset for studying such behaviors.} 
Specifically, we analyzed 12,612 messages to study the evasion behaviors of 94 users. We selected and analyzed these users via a stratified sampling process based on users' flagging frequencies, until thematic saturation was reached. Our analysis shows that users respond to moderations by attempting to bypass them through language modifications in subsequent messages, progressively refining their evasion strategies, resulting in a sustained adversarial cycle. In particular, we observed that, depending on context, users adopt a wide range of bypassing methods, such as multi-line fragmentations, code words, leet speak, alternative spelling strategies, and probing. These techniques exploit limitations of message-level moderation by distributing intent across turns, obfuscating lexical cues, and incrementally probing the system's detection boundaries.
\textbf{Our overall contribution can be summarized as follows.}\label{sec:contributions}
\begin{itemize}
    \item We introduce a \textit{non-intrusive} OCR-based framework for chat data collection from recordings, enabling empirical auditing of moderation systems in gaming environments where data is not available in textual format. In the end, we collected approximately 2 million messages from 105,214 users. 
    \item We develop a saturation-based qualitative thematic analysis approach with LLM-based pre-filtering to extract concrete insights from data at this scale.
    \item We characterize moderation failures by identifying concrete cases of harmful patterns that escaped moderation and also by capturing the evasion strategies used to bypass it.
\end{itemize}

Given the skepticism, this paper takes the first look at Roblox's chat moderation system, only to find how harmful interactions persist (\cref{subsubsec:rq2haconclusions}) and evade detection (\cref{sec:rq2 findings}). We hope these insights and the recommendations (\cref{sec:recommendation}) will help build more robust and context-aware chat moderation systems to protect minors from physical and mental harm.


%% file: Main_document/data_meth.tex
\section{Data Collection}
\label{sec:methodology}
This section describes our data collection process and framework for transforming video recordings into text transcripts.

\subsection{Game Selection}\label{game-selection} 
Since Roblox hosts millions of games, it is infeasible to systematically capture chat data across the entire platform; thus, we focus on popular games. For this study, we selected 4 games appearing in Roblox's ``popular'' section, and have a visible interactive chat window ~\cite{Corporation_2026}. These games were: \textit{Brookhaven RP}, \textit{Adopt Me!}, \textit{Royal High}, and \textit{Berry Avenue RP}. Each game has two age-based settings: (\textit{9+} and \textit{13+}), resulting in eight distinct game-age combinations. Table~\ref{tab:server_size} summarizes the server capacities and total visit counts for each game as of 2024. Here, \textit{server capacity} refers to the maximum number of players that can simultaneously occupy a single game instance or ``server''. While passively observing, we can observe at most the behavior of players co-located on the same server instance at a given time; thus, server size puts an upper bound.

\begin{table}[H]
\centering
\caption{Server size and popularity of games we study.}
\label{tab:server_size}
\begin{tabular}{l c c}
\toprule
 \textbf{Game} & 
\makecell{\textbf{Server Size} \\ \textbf{(Max Players)}} & 
\makecell{\textbf{Total Visits} \\ \textbf{(Billions)}} \\
\midrule
 Brookhaven RP   & 18 & 38.7 \\
 Adopt Me!       & 48 & 34.5 \\
Royal High      & 15 & 9.3  \\
Berry Avenue RP & 30 & 2.0  \\
\bottomrule
\end{tabular}
\end{table}



\begin{figure}[t]
    \centering
    \includegraphics[width=0.48\textwidth]{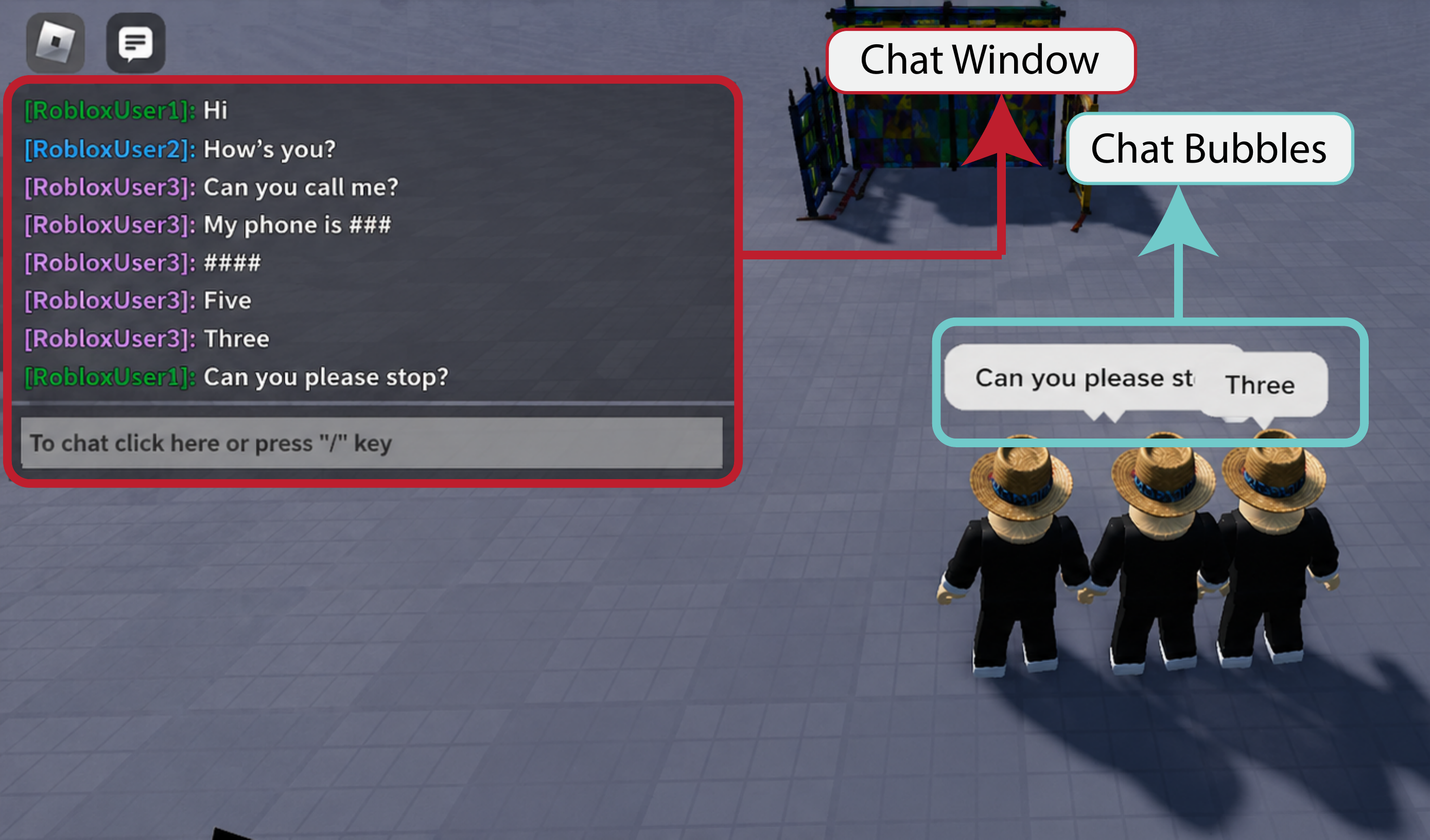}
  \caption{Example of Roblox's graphical user interface with chat bubbles and chat window with dynamic background (anonymized).}
  \label{fig:pipeline1}
\end{figure} 

\subsection{Chat Recording and Frame Deduplication} \label{sec:collect_pipeline}
\noindent
\textbf{Chat Video Recording.}
We record the chat messages as a video stream of screen recording at 60 frames per second (FPS). In Roblox, players primarily interact through live, in-game chat messaging. Roblox displays the chat stream in one of two ways: (1) in a small transparent window anchored to the upper-left corner of the screen, or (2) as a chat bubble above the avatar (Figure~\ref {fig:pipeline1}). Chat window displays chat messages from all players as well as server messages at a fixed screen location in a chronological order. Chat bubbles, by contrast, are spatially tied to moving avatars and display only individual messages, making them less reliable for systematic data collection and transcription. Thus, our recording framework only focuses on the chat window. Since chat messages are moderated before they appear on the screen, we collected the moderated version of the chat stream~\cite{Koneru_2025, Roblox_text_filtering, Roblox_chat_overview}.

Roblox games are hosted across multiple servers (see Table \ref{tab:server_size}) with (variable) maximum user capacity. Servers are typically categorized as: (1) \textit{Public}, owned and managed by the game platform (Roblox) and created dynamically based on user demand~\cite{Roblox_matchmaking, Roblox_scoring, Roblox_attributes_signals, Singh_2025}, or (2) \textit{Private}, owned and controlled by individual users (players). \textit{We recorded chat messages from public servers.}  
Thus, players who enter through the default play flow are algorithmically matched into available public server instances. To prevent servers from preserving inactive sessions, Roblox also removes users who are inactive for 20 minutes~\cite{Roblox_player}. Thus, we used PyAutoGUI~\cite{sweigart2020pyautogui} to automate the mouse and keyboard controls in the game and refresh server connections every hour, allowing us to sample multiple servers over time across recording sessions. For each game and age category, we recorded the chat window continuously for seven consecutive days. Recording ran in approximately 24-hour intervals. In total, we collected 336 hours of chat footage across different time periods in 2024 and 2025.



\noindent
\textbf{Frame Extraction and Deduplication.}
First, each recorded video was converted into a sequence of image frames. Since frame rates might outpace chat rates, the same chat messages may appear across multiple consecutive frames. To deduplicate such frames, we used the structural similarity index (SSIM)-based approach~\cite{villan2019mastering}. SSIM quantifies how similar two images are by comparing luminance, contrast, and structural information~\cite{wang2004image}. This has been shown to perform better than pixel-by-pixel comparisons and has been used in many prior research (e.g,~\cite{eskicioglu1995image}). It produces a similarity score ranging from -1 to 1, where higher values indicate greater visual similarity~\cite{chang2015ssim, mudeng2022prospects}. In our framework, for deduplication, we compared SSIM scores between consecutive frames and discarded the lower-scoring frame if the score was 0.9 or higher. After that, we \textit{crop} the frames to retain only the chat window for transcription.

 \begin{table}[H]
\vspace{5pt}
\centering
\caption{Total Messages in our ground truth dataset.}
\label{tab:randommessages}
\centering
\begin{tabular}{lccc}
\toprule
\textbf{Game} & \textbf{13+} & \textbf{9+} & \textbf{Total per game} \\
\midrule
Adopt Me!        & 2,108   & 1,705  & 3,813  \\
Brookhaven RP    & 980    & 1,062   & 2,042  \\
Royal High       & 442     & 511    & 953   \\
Berry Avenue RP  & 1,962   & 1,961   & 3,923 \\
\midrule
\textbf{Grand Total} & 5,492 & 5,239 & \textbf{10,731} \\
\bottomrule
\end{tabular}%
\end{table}

\subsection{Image-to-Text Conversion}\label{subsec:OCR_meth}

This section describes an optical character recognition (OCR)-based chat transcription method. During our preliminary analysis, we observed that OCR's performance is drastically impaired by the dynamic gameplay background. Thus, drawing inspiration from metamorphic testing~\cite{DBLP:journals/corr/abs-2002-12543}, we created several variations of the original image through careful transformations to assess the OCR model's consistency and confidence.
We first apply grayscale, Gaussian blur, and Otsu binarization along with different polarity settings  (normal/inverted)~\cite{opencvOpenCVImage} to the original image, creating 6 variants of a single image for OCR transcription with PyTesseract\footnote{\url{https://pypi.org/project/pytesseract/}}. Thus, for a given image, we accept the transcription if the OCR's average confidence exceeds 95\%. Otherwise, we first create a new variation by suppressing the background (\cref{background:removal}) to improve the contrast between chat messages and the background. Then, we proceed with the 6 variants of the background-suppressed version and accept the transcription if OCR's average confidence exceeds 95\%.

If OCR's confidence falls below 95\% in this case, we switch from a whole-frame-based approach to a segmentation-based approach. Specifically, we detect text box to localize text regions containing chat lines; localization has shown to improve text recognition performance in complex visual environments~\cite{szeliski2022computer}. 
Note that this line-level OCR operates on smaller, noisier regions, where relying on near-perfect confidence would discard many usable transcriptions and reduce overall recall. Thus, the specific confidence threshold (median confidence $\ge 74\%$, mean confidence $\ge \%70$) for this was determined through manual investigation of a random subset. After transcription, we deduplicate text messages based on a similarity threshold of 85\%, i.e., if two strings are more than 85\% similar, they are removed from the final transcription. Again, we determine this threshold by manual observation. After that, we anonymize (\cref{final:dataset}) the dataset before analyzing it.


\subsection{Transcription Framework Performance}\label{subsec:performance:validation}

To measure the performance of our OCR-based video transcription framework, we first manually create a ground truth dataset from the video, then compare its performance against this dataset.

\noindent
\textbf{Ground truth Transcription Dataset Creation.} To create our ground truth dataset, we randomly select 20-minute videos across all the games to a total recording duration of two hours and manually transcribe them. Table \ref{tab:randommessages} shows a summary of it.


\noindent
\textbf{Comparison Result.} Our performance comparison mainly answers three questions: (1) \textit{Ablation study 1:} how would our framework perform if it were only implemented with original image variations? (2) \textit{Ablation study 2:} How would our framework perform if it were only implemented with variations of the background-suppressed version? and (3) \textit{Overall performance:} how does our final OCR-based transcription framework perform in comparison with the ground truth? \underline{Metrics:} For performance comparison we use two metrics: \textit{ Recall} and \textit{Average Matched Similarity (AMS)}. Recall represents the proportion of chats successfully recovered by the OCR-based system in comparison with the ground truth, while \textit{AMS} measures how closely the recovered messages match the corresponding ones from the ground truth set (Details in~\cref{metrics:evaluation}). Results are summarized in~\cref{tab:ocrstats}. 

\noindent
\underline{Ablation study 1:} We found that by only using the original image variants, we can retain $71\%$ of the chats with an average matched similarity of $96\%$ in comparison with the ground truth set. This indicates that while the OCR system produces highly accurate text when matches are found, a portion of the ground-truth lines remain undetected due to missed extractions. The low match rate is primarily due to the overlay effect of background changes. 

\noindent
\underline{Ablation study 2:} Interestingly, OCR-based transcription with the variants of our background-suppressed version alone could retain 80\%, which is $\sim$9\% more compared to original image variants, but surprisingly, the overall similarity score dropped to 90.89\%. 

\noindent
\underline{Overall Performance:} As shown in~\cref{tab:ocrstats}, the performance of our final OCR-based transcription framework is much higher than the variants of original image or background-suppressed methods, individually. Specifically, it could retain 11\% more chat messages (91.76\% of the ground truth set) with a higher similarity score (97.44\%).

\vspace{2mm}


\begin{table}[h]
\caption{Comparison of OCR text recovery across original image variants, background-suppressed image variants, and our transcription framework with ground truth chat transcriptions. Here, Recall denotes the percentage of retained chat messages relative to the ground truth set, and AMS stands for average matched similarity, which measures the similarity between the retained chats and their corresponding ground truth.}
\label{tab:ocrstats}
\centering
\begin{tabular}{lcc}
\toprule
\textbf{Type (Variation Counts)} & \textbf{Recall} & \textbf{AMS} \\
\midrule
Original Image Variants (6)       & 71.09\% & 96.09\% \\
Background-Suppressed Variants (6)      & 80.27\% & 90.89\% \\
\textbf{Our Framework} & \textbf{91.76\%} & \textbf{97.44\%} \\
\bottomrule
\end{tabular}
\end{table}

\begin{table}[h]
\centering
\caption{Overall statistics of our chat dataset across 4 different games. Here, ``Affected Lines'' refers to chat messages affected by moderation (i.e., parts of them replaced with ``\#\#\#''), ``Masked Contents'' refers to the total number of contents (e.g., words) in chat messages that were masked.}
\label{tab:game_hash_profile}
\small
\begin{tabular}{lccccr}
\toprule
Game & Total & Total & Affected & Masked & Avg. Span\\
     & Users & Lines & Lines  & Contents  & Len.\\
\midrule
AdoptMe      & 16,813 & 201,252 & 5,956  & 10,926 & 7.46\\
BerryAve     & 25,290 & 504,891 & 31,986 & 61,180 & 5.93\\
Brookhaven   & 46,425 & 900,881 & 37,550 & 58,552 & 6.42\\
RoyaleHigh   & 16,686 & 333,981 & 13,231 & 22,127 & 7.42\\
\midrule
\textbf{Total}   & \textbf{105,214} & \textbf{2,040,839} & \textbf{88,273} & \textbf{152,785} & \\
\bottomrule
\end{tabular}
\end{table}

\subsection{Final Dataset}\label{final:dataset}

After processing all the chat recordings with our transcription framework, we produced a corpus of $\sim 2~million$ chat messages. A summary of this corpus is presented in Table~\ref{tab:game_hash_profile}.  

\noindent
\textbf{Anonymizing the dataset.}
After transcription, we anonymized usernames and redacted any personally identifiable information (PII) that users may have intentionally or inadvertently shared (details in~\cref{anonymization:dataset}). Although our analysis is performed using local models, we enforce anonymization to ensure ethical handling of users' data and to prevent potential exposure through intermediate artifacts such as logs, outputs, or illustrative examples. These steps ensure that our analysis remains both methodologically robust and aligned with ethical standards.

\section{RQ1 (Escaped Categories): Method} \label{sec:LLM_approach}



Here we detail how we identified unsafe messages that bypassed Roblox's moderation. Given the scale of our dataset, it is prohibitively expensive to perform rigorous qualitative content analysis. Inspired by Thomas~\textit{et al.}~\cite{DBLP:conf/sp/ThomasKTMVTBFEB25} \textit{pre-filtering design pattern} for using LLMs to filter out safe content and pass only unsafe content for human review, we design a two-stage framework for our \textit{post hoc} analysis as well. Next, we discuss how we designed this framework.

\subsection{Messages to Conversations}
Given a chat message, both humans and LLMs would require the \textit{context} under which the classification is made. This is because, given the surrounding conversation, a given message can be either benign or unsafe (e.g., grooming). Thus, to support analysis at the conversation level, we organized the chat messages into 50-line conversations, yielding a total of 38,856 chat conversations. We selected this cluster size to preserve conversational context within a single interaction thread, including turn-taking, i.e., the sequential exchange of messages between users, intent progression, and response dependencies, while maintaining input lengths that are manageable for both LLM-based analysis and manual review. This design also mitigates long-context limitations such as the ``Lost-in-the-Middle'' (LiM) effect, where models tend to underutilize information positioned in the middle of long sequences~\cite{veseli2025positional}.

\begin{table}[H]
\caption{Labels generated with \textit{gpt-oss-120B} with few-shot prompting}
\label{tab:gpt-oss-stats}
\centering
\begin{tabular}{lc}
\toprule
\textbf{Label} & \textbf{Count} \\
\midrule
Absolutely UNSAFE        & 5,208  \\
Possibly SAFE       & 5,734  \\
Absolutely SAFE  & 24,834  \\
Possibly SAFE & 1,081 \\

\bottomrule
\end{tabular}
\end{table}

\subsection{Pre-filtering and Thematic Analysis} \label{sec:thematic_analysis}

\noindent
\textbf{Pre-filtering with LLMs.} For pre-filtering safe messages on the full dataset, we used \textit{gpt-oss-120B} with few-shot prompting. The selection process was guided with a manually curated ground truth sample of 2,000 conversations, where~\textit{gpt-oss-120B} with few-shot prompting achieved 44.13\% F1 score with 31.39\% precision and 71.43\% recall (details in~\cref{subsec:llm:selection}). Here, low precision would eventually lead to more manual effort; thus, we adopt a saturation-based technique. However, given saturation-based analysis and the fact that we are missing at least 1 our of 4 unsafe messages, these results indicate the lower-bound estimate of the prevalence of unsafe interactions.
The results after the filtration are shown in~\cref{tab:gpt-oss-stats}. For the sake of analysis feasibility, for unsafe message analysis, we only consider the messages labeled as unsafe with high confidence (``Absolutely UNSAFE''). Next, for these messages, we extracted category-specific keywords (e.g., grooming'', sexual content'', bullying'', harassment'', profanity'', and attempts to move conversations ``off-platform'') from the model-generated explanations associated with each conversation and categorized the conversations based on these keywords to enable category-specific thematic analysis.

\noindent
\textbf{Saturation-based Thematic Analysis.} The unsafe message corpus, organized into categories, serves as the basis for our qualitative analysis. We treat each category as an independent analytical unit and conduct an inductive thematic analysis within each category. Specifically, we perform the analysis iteratively by sampling conversations across strata, where each stratum corresponds to a game (by age group) within a category, ensuring coverage of diverse interaction contexts. In each iteration, we randomly select one conversation per stratum and code it based on themes, which are iteratively refined. We define thematic saturation for a category as the point at which no new violation patterns emerge for 
5 iterations, and declare saturation when successive iterations across strata yield no additional patterns~\cite{Mpofu31122026}. 

\begin{table}[h]
\caption{Distribution of unsafe content categories across the full dataset and the reviewed subset (until saturation).} 


\label{tab:unsafe-combined}
\footnotesize
\centering
\begin{tabular}{lcccc}
\toprule
\textbf{Label} & \textbf{Total Conv.} & \textbf{Total Reviewed Conv.}  & \textbf{TP/FP}\\
 &  & \textbf{(Messages)}  & \\
\midrule
Profanity & 1,425 & 40 (1,600) & 32/8  \\
Bullying \& Harassment & 1,350  & 60 (2,550) & 51/9  \\
Grooming & 1,255 & 80  (2,666)  & 54/26   \\
Sexual Content & 1,231 & 80 (1,950)  & 39/41  \\
Threats / Violence & 923 & 60 (2,450)  & 49/11  \\
Redirecting off-platform & 40   & 37 (1,850) & 37/3  \\
Hate Speech & 273 & 40 (1,450)   & 29/11  \\
Self-harm Related Content & 192  & 120 (700)  & 14/6   \\
Request for PII & 91 & 20 (850) & 17/3   \\
\midrule
\textbf{Total} & \textbf{7,254} & \textbf{440 (16,066)}  & \textbf{322/118}   \\
\bottomrule
\end{tabular}
\end{table}




In Table~\ref{tab:unsafe-combined}, we show the distribution of unsafe categories based on how the model characterizes the prevalence of different types of harmful behaviors across the corpus. These ``Labels'' are inspired by Roblox community standards~\cite{robloxcomm}, where they specify their tolerance policy for the platform. It is important to note that these counts are not mutually exclusive; rather, they represent the frequency of specific unsafe behaviors extracted from LLM's reasoning. For example, a single conversation may contribute to multiple categories (e.g., grooming co-occurring with requests for personal information or off-platform migration), leading to overlapping counts across categories. 
Table~\ref{tab:unsafe-combined} also reports the number of instances reviewed during our saturation-based thematic analysis, along with those identified as false positives. Consequently, the absolute counts of unsafe messages in Table~\ref{tab:unsafe-combined} should not be interpreted as exact estimates; instead, they are more appropriately used to compare the relative prominence of different categories of unsafe content.

\section{RQ1 (Escaped Categories): Findings} \label{subsubsec:rq2haconclusions}

The following subsections summarize each category of chat safety violations with representative examples from the annotated corpus. Signals for unsafe messages are highlighted in \textcolor{red}{red}. Red texts in square braces (\textcolor{red}{[text]}) indicate either that the content is anonymized for privacy or rephrased for explicitness.

\subsection{Grooming} 


Grooming was the largest category in the reviewed corpus, in part because it often combined several harmful strategies in a single conversation. Prior research on child grooming defines it as \textit{a predatory behavior where users attempt to build trust, move communication into less moderated spaces, or normalize sexual interaction with a minor} ~\cite{craven2006sexual}. These harmful strategies that enable grooming on such patterns include, but are not limited to, manipulative behavior, requests for usernames or off-platform channels, questions about age or gender, sexualized prompts, intimidation, emotional targeting, and the disclosure or solicitation of identifying information. 

\noindent
\textbf{\colorbox{black!20}{\#1:} Persuading for in-person meetings.} In the following case, the interaction moves from a desire to see another player toward location disclosure and an in-person meeting. While not harmful independently, taken collectively, these patterns signal a trajectory that may expose children to exploitation. A prime example is the California kidnap case~\cite{Tan_2025}, which indicates that sharing a physical address and arranging to meet strangers from Roblox can escalate to real-world harm, underscoring the risks of such interactions and the need for early intervention~\cite{whittle2013review}.

\begin{tcolorbox}[colback=gray!7!white, colframe=black,
  boxrule=1pt,
  left=1mm,
  right=1mm,
  top=1mm,
  bottom=1mm,
  fontupper=\ttfamily\small
]
\footnotesize
\textbf{Grooming Intended to Disclose Location and In-Person Meeting}\par\smallskip

\textbf{\texttt{User1:}} \texttt{``I want to see you''} $\rightarrow$
\textbf{\texttt{User2:}} \texttt{``me too i’m in \textcolor{red}{\textbf{[728]}} with a \textcolor{red}{\textbf{[blue toyota and white ford]}}''} $\rightarrow$
\textbf{\texttt{User2:}} \texttt{``well the \textcolor{red}{\textbf{[white ford]}} is not there rn''} $\rightarrow$
\textbf{\texttt{User2:}} \texttt{``omg what add''} $\rightarrow$
\textbf{\texttt{User2:}} \texttt{``dress''} $\rightarrow$
\textbf{\texttt{User1:}} \texttt{``Let’s grind''} $\rightarrow$
\textbf{\texttt{User2:}} \texttt{``what address''} $\rightarrow$
\textbf{\texttt{User2:}} \texttt{\#\#\#\#\#\#\#} $\rightarrow$
\textbf{\texttt{User1:}} \texttt{``Ok my mom doesn’t tell anyone but my guests''} $\rightarrow$
\textbf{\texttt{User2:}} \texttt{``i?’m so glad we’re close''} $\rightarrow$
\textbf{\texttt{User1:}} \texttt{``like iwanna meet u today''} ...
\end{tcolorbox}

Realization of potential harms due to personal detail leakage can cause emotional trauma originating from fear and anxiety~\cite{Ramadhani_Khodari_Ulfiah_Rosydawati_2025}. We observed cases when users realized the risks of sharing such info afterward and panicked about their safety, as shown below.

\begin{tcolorbox}[colback=gray!7!white, colframe=black,
  boxrule=1pt,
  left=1mm,
  right=1mm,
  top=1mm,
  bottom=1mm,
  fontupper=\ttfamily\small
]
\footnotesize
\textbf{Panic After Sharing Location Infortion}\par\smallskip

\textbf{\texttt{User1:}} \textcolor{red}{\textbf{\texttt{``IM FREAKING OUT''}}} $\rightarrow$
\textbf{\texttt{User2:}} \texttt{``User3 someone texted User1''} $\rightarrow$
\textbf{\texttt{User2:}} \texttt``{and said they know where she lives''} $\rightarrow$
\textbf{\texttt{User3:}} \texttt{``WAIT WHAT''} $\rightarrow$
\textbf{\texttt{User3:}} \texttt{``BLOCK IT''} $\rightarrow$
\textbf{\texttt{User3:}} \texttt{``NOW''} $\rightarrow$
\textbf{\texttt{User2:}} \texttt{``Ask them to say your exact location''} $\rightarrow$
\textbf{\texttt{User1:}} \texttt{``I DID''} $\rightarrow$
\textbf{\texttt{User1:}} \texttt{``THEY KNOW''} $\rightarrow$
\textbf{\texttt{User3:}} \texttt{``THEY KNOW UR EXACT LOCATION?!''} $\rightarrow$
\textbf{\texttt{User3:}} \texttt{``HURRY AND BLOCK THE NUMBER RN''} $\rightarrow$
\textbf{\texttt{User4:}} \textbf{\textcolor{red}{\texttt{``and havent u herd or gamecharlie1''}}}\par
\end{tcolorbox}

After investigation, we found that \texttt{gamecharlie1} is an urban legend within the Roblox community about an 11-year-old girl who was groomed and kidnapped after sharing personal information in Roblox~\cite{Fandom_2021}. This story is often used as a cautionary tale.


\noindent
\textbf{\colorbox{black!20}{\#2:} Persuading for sexual exploitation.} Another prevalent case of grooming is persuading seemingly underage children for sexual exploitation either online or in-person, as shown below.

\begin{tcolorbox}[colback=gray!7!white, colframe=black,
  boxrule=1pt,
  left=1mm,
  right=1mm,
  top=1mm,
  bottom=1mm,
  fontupper=\ttfamily\small
]
\footnotesize
\textbf{Grooming Intended to Disclosure Age and Sexual Insinuations}\par\smallskip

\textbf{\texttt{User3:}} \texttt{``how old r u?''} $\rightarrow$
\textbf{\texttt{User4:}} \texttt{\textcolor{red}{\textbf{``nine''}}} $\rightarrow$
\textbf{\texttt{User3:}} \texttt{``are you into anything?''} $\rightarrow$
\textbf{\texttt{User3:}} \texttt{``do you want to like do something''} $\rightarrow$
\textbf{\texttt{User3:}} \texttt{``like this''} $\rightarrow$
\textbf{\texttt{User4:}} \texttt{``Sure''} $\rightarrow$
\textbf{\texttt{User4:}} \texttt{``I'm be back 1. A min''}  $\rightarrow$
\textbf{\texttt{User3:}} \texttt{\textcolor{red}{``abc if you want to have s \textbf{(suggestive for Sex)''}}}\par
\end{tcolorbox}

\begin{tcolorbox}[colback=gray!7!white, colframe=black,
  boxrule=1pt,
  left=1mm,
  right=1mm,
  top=1mm,
  bottom=1mm,
  fontupper=\ttfamily\small
]
\footnotesize
\textbf{Persuasion for Online Sexual Engagement}\par\smallskip

\textbf{\texttt{User5:}} \texttt{``Party ??''} $\rightarrow$
\textbf{\texttt{User6:}} \texttt{``Adult?''} $\rightarrow$
\textbf{\texttt{User5:}} \texttt{yes!} $\rightarrow$
\textbf{\texttt{User6:}} \texttt{\textcolor{red}{``both of us''}}$\rightarrow$
\textbf{\texttt{User6:}} \texttt{``get it''} $\rightarrow$
\textbf{\texttt{User6:}} \texttt{\textcolor{red}{``create a babyy or babys''}} $\rightarrow$
\textbf{\texttt{User5:}} \texttt{\textbf{\textcolor{red}{``Come to my home ok''}}} $\rightarrow$
\textbf{\texttt{User3:}} {onnnhhhhhhhhhhhnnn}\par
\end{tcolorbox}

\noindent
\textbf{\colorbox{black!20}{\#3:} Persuading to share personal explicit images.} We also observed chat conversations where one user tries to persuade the other (potentially underage) to share personal, explicit images. At times, these chats can turn into transactional coercions too. In the following example, \textit{User1} pressures \textit{User2} to share explicit images. The request begins with a seemingly ordinary request for Korblox -- a high-cost, elite, and popular avatar bundle series~\cite{Deathwalker_2026}. The request begins with a seemingly ordinary request for korblox, but the conversation shifts towards an intimate image request and bargaining over the type of image expected. The rest of the content in this conversation is not displayed on purpose because it is explicit.

\begin{tcolorbox}[colback=gray!7!white, colframe=black,
  boxrule=1pt,
  left=1mm,
  right=1mm,
  top=1mm,
  bottom=1mm,
  fontupper=\ttfamily\small
]
\footnotesize
\textbf{Transactional Coercion for Sharing Explicit Images}\par\smallskip

\textbf{\texttt{User1:}} \texttt{``buy me korblox''} $\rightarrow$
\textbf{\texttt{User2:}} \texttt{``\textcolor{red}{ok send me}''} ... \texttt{``\textcolor{red}{\textbf{a nak}}''} $\rightarrow$
\textbf{\texttt{User1:}} \texttt{``buy me korblox for a face pic''} $\rightarrow$
\textbf{\texttt{User2:}} \textcolor{red}{Negotiates and requests [\textbf{explicit content}]}
\end{tcolorbox}
Sharing such sensitive content with strangers can lead to exploitation, extortion, and long-term harm. 
In such cases, early interventions can help prevent harm and mitigate lasting psychological distress in children~\cite{Sravanti2025}.


\noindent
\textbf{\colorbox{black!20}{\#4:} Persuading to share off-platform accounts.}
Other instances of grooming we observed are to shift communication to an external platform (e.g., \textit{TikTok}) along with mild flirting (e.g., calling \emph{``sweetie''}) to gain trust, thereby moving the conversation beyond the moderated environment and reducing oversight.

\begin{tcolorbox}[colback=gray!7!white, colframe=black,
  boxrule=1pt,
  left=1mm,
  right=1mm,
  top=1mm,
  bottom=1mm,
  fontupper=\ttfamily\small
]
\footnotesize
\textbf{Grooming-Oriented Platform Migration}\par\smallskip

\textbf{\texttt{User2:}} \texttt{``What other apps do you have that you can text''} $\rightarrow$
\textbf{\texttt{User1:}} \texttt{``what do u have''} $\rightarrow$
\textbf{\texttt{User2:}} \texttt{\textcolor{red}{\textbf{``Tt'' (Shortform of Tiktok)}}}\par

\end{tcolorbox}

\begin{tcolorbox}[colback=gray!7!white, colframe=black,
  boxrule=1pt,
  left=1mm,
  right=1mm,
  top=1mm,
  bottom=1mm,
  fontupper=\ttfamily\small
]
\footnotesize
\textbf{Grooming-Oriented flirt and Platform Migration}\par\smallskip

\textbf{\texttt{User1:}} \texttt{``That’s a compliment sweetie''} $\rightarrow$
\textbf{\texttt{User2:}} \texttt{``miss''} $\rightarrow$
\textbf{\texttt{User2:}} \texttt{``E dates''} $\rightarrow$
\textbf{\texttt{User2:}} \texttt{``here...''} $\rightarrow$
\textbf{\texttt{User1:}} \texttt{\textcolor{red}{``how do you know my first name''}} $\rightarrow$
\textbf{\texttt{User2:}} \texttt{``IKR''} $\rightarrow$
\textbf{\texttt{User1:}} \texttt{\textcolor{red}{\textbf{``Wanna resolve it over vc?'' (short for Video call/chat)}}} $\rightarrow$
\textbf{\texttt{User2:}} \texttt{``yes babe''} ...
\end{tcolorbox}

\subsection{Bullying, Harassment, and Hate Speech}
This category captures targeted insults, intimidation, ridicule, repeated antagonism, and discriminatory abuse directed at another user or identity group. Although bullying and harassment, and hate speech were labeled separately during annotation, we discuss them together here because they frequently appeared in the same context. Recent work further supports this approach, arguing that intent is inexorably intertwined and involves multiple dimensions~\cite{Wang_Rajtmajer_2025, Garcia_70029}. In the reviewed conversations, discriminatory language was often used as a harassment tactic, using race, gender identity, sexuality, appearance, or social status to degrade or provoke another player. As a result, separating hate speech from bullying and harassment sometimes obscured how these harms operated together in chat. Next, we discuss some examples based on their types.

\noindent
\textbf{\colorbox{black!20}{\#5:} Use of profane abuse for bullying and harassments.} The most prevalent case of bullying and harrasment involves the use of profane and deregatory language directed at other users as shown in the following example.

\begin{tcolorbox}[colback=gray!7!white, colframe=black,
  boxrule=1pt,
  left=1mm,
  right=1mm,
  top=1mm,
  bottom=1mm,
  fontupper=\ttfamily\small
]
\footnotesize
\textbf{Use of profane language for bullying}\par\smallskip

\textbf{\texttt{User1:}} \texttt{``So Don't F4 Tell''} $\rightarrow$
\textbf{\texttt{User1:}} \texttt{``first off stop acting ghetto''} $\rightarrow$
\textbf{\texttt{User2:}} \texttt{``GhEtTo''} $\rightarrow$
\textbf{\texttt{User1:}} \texttt{``Ur hella embarrassing''} $\rightarrow$
\textbf{\texttt{User1:}} \texttt{\textcolor{red}{``Btc'' (Alternate for Bitch)}} $\rightarrow$
\textbf{\texttt{User1:}} \texttt{\textcolor{red}{``Look at your ugly assi'' (``ass'')}} $\rightarrow$
\textbf{\texttt{User2:}} \texttt{Newer} $\rightarrow$
\textbf{\texttt{User1:}} \texttt{``Sthu''} ...
\end{tcolorbox}

Our analysis shows that Roblox's moderation system usually masks common profane words. Thus, users would oftentimes use coded language to bypass the moderation. We discuss the nature of profane language itself in~\cref{sec:profanity} and the common techniques used to bypass the moderations in~\cref{sec:rq2 findings}. 

\noindent
\textbf{\colorbox{black!20}{\#6:} Harassments through racial remarks.} Another prevalent case of harassment is the use of racial remarks. The following example shows racialized harassment, where one user interprets another user's ``monkey'' comment as targeting their Black identity and explicitly identifies the exchange as racist. Although Roblox masked part of the final message, the surrounding context still makes the racialized insult recoverable. Collectively, these examples show that harassment often operates through escalation, with discriminatory language and identity-based degradation (in the form of hate speech) used to intensify interpersonal conflict.

\begin{tcolorbox}[colback=gray!7!white, colframe=black,
  boxrule=1pt,
  left=1mm,
  right=1mm,
  top=1mm,
  bottom=1mm,
  fontupper=\ttfamily\small
]
\footnotesize
\textbf{Social Degradation}\par\smallskip

\textbf{\texttt{User3:}} Racism runs both ways $\rightarrow$
\textbf{\texttt{User4:}} \textcolor{red}{``Cuz black figures always thinking about seafood''} \#\#\# $\rightarrow$
\textbf{\texttt{User4:}} Look atu $\rightarrow$
\textbf{\texttt{User3:}} Yes $\rightarrow$
\textbf{\texttt{User4:}} AHEN $\rightarrow$
\textbf{\texttt{User3:}} Look at u lol $\rightarrow$
\textbf{\texttt{User4:}} ever said such things $\rightarrow$
\textbf{\texttt{User3:}} Ong $\rightarrow$
\textbf{\texttt{User4:}} \textcolor{red}{``This monkey''} $\rightarrow$
\textbf{\texttt{User4:}} Is \#\#\#\#\#\# $\rightarrow$
\textbf{\texttt{User3:}} That’s racist $\rightarrow$
\textbf{\texttt{User4:}} Cuz everyone is monkeys $\rightarrow$
\textbf{\texttt{User3:}} \textcolor{red}{``Ima monkey bc I’m black''} $\rightarrow$
\textbf{\texttt{User3:}} That’s how u acting $\rightarrow$
\textbf{\texttt{User4:}} \textcolor{red}{``U can be white while being a \#\#\#\#\#\# \# \# \#\#\#\#''}
\end{tcolorbox}

Note that, in the final message from \textit{User4}, moderation is applied at the phrase level rather than at the individual word level. This suggests that Roblox’s moderation is not purely keyword-based, but instead leverages AI to capture contextual meaning, as described in their official documentation~\cite{Roblox_bert}. However, the observed outcome highlights inconsistencies in such AI-based systems, as the majority of the harmful exchanges remained unfiltered.

\noindent
\textbf{\colorbox{black!20}{\#7:} Hate speech based on race and identity.} Hate speech refers to expressions that attack, demean, or incite hostility or violence against an individual or group based on protected characteristics such as race, ethnicity, religion, gender, sexual orientation, or disability. The following are some examples of hate speech we observed based on race and sexual orientation.

\begin{tcolorbox}[colback=gray!7!white, colframe=black,
  boxrule=1pt,
  left=1mm,
  right=1mm,
  top=1mm,
  bottom=1mm,
  fontupper=\ttfamily\small
]
\footnotesize
\textbf{Explicit Racist Hate Speech and Slur}\par\smallskip
\textbf{\texttt{User3:}} \textcolor{red}{\textbf{``NILGG ERS = SLAVES''}} $\rightarrow$
\textbf{\texttt{User3:}} ``\#H\#\# ck lives, D,.ONT. matter.'' $\rightarrow$
\textbf{\texttt{User3:}} ``\#\#\# LIVE ES MAntER‘(ekcept niggers)'' $\rightarrow$
\textbf{\texttt{User3:}} ``\#\#\# about you quit pick ing cot ton and pick uP'' $\rightarrow$
\textbf{\texttt{User3:}} ``wedontsellniggas*''
\end{tcolorbox}

\begin{tcolorbox}[colback=gray!7!white, colframe=black,
  boxrule=1pt,
  left=1mm,
  right=1mm,
  top=1mm,
  bottom=1mm,
  fontupper=\ttfamily\small
]
\footnotesize
\textbf{Hate Speech based on Sexual Orientation}\par\smallskip

\textbf{\texttt{User4:}} \textcolor{red}{\textbf{``Ru being home o phobic''}} $\rightarrow$
\textbf{\texttt{User4:}} \textcolor{red}{\textbf{`Is it bc 'm trans ... Ru telling her to die''}} $\rightarrow$
\textbf{\texttt{User5:}} ``Ur not the government'' $\rightarrow$
\textbf{\texttt{User4:}} ``niether r yall'' $\rightarrow$
\textbf{\texttt{User4:}} ``Atleast we aren’t blonde basic and skinny'' $\rightarrow$
\textbf{\texttt{User5:}} ``Little miss is having a tantrum''
\end{tcolorbox}

\subsection{Sexual Content}
This category captures messages involving sexualized language, propositions, suggestive roleplay, flirtatious escalation, and requests for intimate content. Sexual content often appeared in partial words, euphemistic phrasing, fragmented spelling, or context-dependent innuendo. In some cases, users described simulated sexual contact, unwanted pregnancy, etc. These conversations included extended adult-themed content, raising concerns about children's exposure to age-inappropriate narratives. Although we exclude these examples due to their disturbing nature, these examples underscore the importance of identifying sexualized content as a developmental safety concern, as adolescent health research links early sexual attitudes, dating violence, sexting, and sexual risk behaviors to broader reproductive health risks and sexual normalization~\cite{Ruvalcaba_Mercer, Kabiru, Sravanti2025}.

\noindent
\textbf{\colorbox{black!20}{\#8:} Soliciting Sex.} A prevalent pattern involves users explicitly or implicitly requesting sexual content or engaging others in sexually suggestive exchanges. Some of the example phrases used for such solicitation are shown below.

\begin{tcolorbox}[colback=gray!7!white, colframe=black,
  boxrule=1pt,
  left=1mm,
  right=1mm,
  top=1mm,
  bottom=1mm,
  fontupper=\ttfamily\small
]
\footnotesize
\textbf{Solicitation for Sex (Example Phrases)}\par\smallskip
\textbf{\texttt{Phrase 1:}}  \textcolor{red}{``Can you give me a H.J'' (\textit{Hand-Job})}\par
\textbf{\texttt{Phrase 2:}} \textcolor{red}{``IM HURNYY (\textit{Horny})''}\par
\textbf{\texttt{Phrase 3:}} \textcolor{red}{``anyone know of any good sugar daddies around''}\par
\end{tcolorbox}

\noindent
\textbf{\colorbox{black!20}{\#9:} Sexting and roleplay.} Sexting~\cite{doring2014consensual} is defined as the exchange of sexually explicit or suggestive messages, often in a conversational or roleplay context. We observe that such interactions begin as role-play and gradually escalate into more explicit exchanges. Examples like the ones below indicate the challenges for real-time moderation to detect such cases, as they often unfold incrementally across messages and rely on implicit context.

\begin{tcolorbox}[colback=gray!7!white, colframe=black,
  boxrule=1pt,
  left=1mm,
  right=1mm,
  top=1mm,
  bottom=1mm,
  fontupper=\ttfamily\small
]
\footnotesize
\textbf{Roleplay using implicit language}\par\smallskip
\textbf{\texttt{User1:}} \textcolor{red}{``Do you want me to do it hotter?''} $\rightarrow$
\textbf{\texttt{User2:}} ``Oh yeah'' $\rightarrow$
\textbf{\texttt{User1:}} \textcolor{red}{``Do you want me to put it in there?''}\par
\end{tcolorbox}

\begin{tcolorbox}[colback=gray!7!white, colframe=black,
  boxrule=1pt,
  left=1mm,
  right=1mm,
  top=1mm,
  bottom=1mm,
  fontupper=\ttfamily\small
]
\footnotesize
\textbf{Roleplay using explicit language}\par\smallskip
\textbf{\texttt{User1:}} \textcolor{red}{``in mouth''} $\rightarrow$
\textbf{\texttt{User1:}} \textcolor{red}{``I \textbf{[Explicit content]}''}\par
\end{tcolorbox}

\noindent
\textbf{\colorbox{black!20}{\#10:} Objectifying with explicit language.} Another prevalent pattern involves users reducing others to sexual attributes or body parts using explicit language as shown below. Such objectification exposes minors to inappropriate sexual content, normalizes harmful behavior, and increases their vulnerability to grooming and exploitation.

\begin{tcolorbox}[colback=gray!7!white, colframe=black,
  boxrule=1pt,
  left=1mm,
  right=1mm,
  top=1mm,
  bottom=1mm,
  fontupper=\ttfamily\small
]
\footnotesize

\textbf{Objectifying with explicit language}\par\smallskip
\textbf{\texttt{User1:}} \textcolor{red}{``bigassnegro''} $\rightarrow$
\textbf{\texttt{User2:}} ``a pretty one tonight'' $\rightarrow$
\textbf{\texttt{User2:}} ``ima girl'' $\rightarrow$
\textbf{\texttt{User2:}} \textcolor{red}{``if u were a king would u make me ur queen''} $\rightarrow$
\textbf{\texttt{User3:}} \textcolor{red}{``bigblacktittie''} $\rightarrow$
\textbf{\texttt{User2:}} ``soo ami pretty'' $\rightarrow$
\textbf{\texttt{User2:}} \textcolor{red}{``do u wanna get busy''}
\end{tcolorbox}

\noindent
\textbf{\colorbox{black!20}{\#11:} Sexualized rumors about other entities.} We also observed several instances of users spreading or amplifying sexualized rumors about peers, public figures, or authority figures, often naming individuals and attaching explicit or suggestive claims. The following example shows such an instance involving teachers.
\begin{tcolorbox}[colback=gray!7!white, colframe=black,
  boxrule=1pt,
  left=1mm,
  right=1mm,
  top=1mm,
  bottom=1mm,
  fontupper=\ttfamily\small
]
\footnotesize
\textbf{Sexualized Rumor Involving an Authority Figure}\par\smallskip
\textbf{\texttt{User1:}} \texttt{``I heard \textcolor{red}{[name]} had \#\#\# with the teacher''} $\rightarrow$
\textbf{\texttt{User2:}} \texttt{``NOT THE SCINCE TEACHER''} ... \texttt{``ACTUALLY THAT MAKES SENSE.''} $\rightarrow$
\textbf{\texttt{User1:}} \texttt{``HE RLLY} \textcolor{red}{[\textbf{explicit content}]} \texttt{SO SHE COULD GET AN A.''}\par
\end{tcolorbox}

\subsection{Threats, Violence, and Self-Related Harm}
Threats, violence, and self-related harm captures messages that describe, encourage, or fantasize about physical harm toward another person or character. In the reviewed conversations, these included direct threats, descriptions of severe physical injury, encouragement of violence, and disturbing references to forced or self-directed harm. Although some cases appear as exaggerated roleplay or joking escalation, they remain visible in chat and consequently expose others, especially minors, to violent imagery and harmful language. Prior research suggests that this exposure can affect emotional development, increasing distress while contributing to desensitization of violent behavior towards others~\cite{mrug2016emotional}. 

\noindent
\textbf{\colorbox{black!20}{\#12:} Threats and violence.}
We show two representative examples of threats and violence below.

\begin{tcolorbox}[colback=gray!7!white, colframe=black,
  boxrule=1pt,
  left=1mm,
  right=1mm,
  top=1mm,
  bottom=1mm,
  fontupper=\ttfamily\small
]
\footnotesize
\textbf{Forced-Harm Scenario in Roleplay}\par\smallskip

\textbf{\texttt{User1:}} \texttt{``kind of created a demon inside \textcolor{red}{[name]}''} $\rightarrow$
\textbf{\texttt{User2:}} \texttt{``WHY WOULD YOU DO THAT''} $\rightarrow$
\textbf{\texttt{User1:}} \texttt{``\textcolor{red}{[name]} a demon''} $\rightarrow$
\textbf{\texttt{User2:}} \texttt{``Yea uhm''} $\rightarrow$
\textbf{\texttt{User1:}} \textcolor{red}{``made her jump of the roof''} $\rightarrow$
\textbf{\texttt{User1:}} \textcolor{red}{``Of a random house''}
\end{tcolorbox}

\begin{tcolorbox}[colback=gray!7!white, colframe=black,
  boxrule=1pt,
  left=1mm,
  right=1mm,
  top=1mm,
  bottom=1mm,
  fontupper=\ttfamily\small
]
\footnotesize
\textbf{Escalating Threats of Physical Violence}\par\smallskip

\textbf{\texttt{User3:}} 
\textbf{\texttt{User3:}} \texttt{\textcolor{red}{``WILL RIP HIS VOCAL CORDS OUT''}} $\rightarrow$
\textbf{\texttt{User3:}} \texttt{\textcolor{red}{``OR HIS LUNGS''}} $\rightarrow$
\textbf{\texttt{User3:}} \texttt{``OR MAYBEEEEE''} $\rightarrow$
\textbf{\texttt{User4:}} \texttt{``just kick him''} $\rightarrow$
\textbf{\texttt{User4:}} \texttt{``many times''} $\rightarrow$
\textbf{\texttt{User4:}} \texttt{\textcolor{red}{``to where he is near the brink of death''}}\par
\end{tcolorbox}

The first describes making another character jump from a roof, blending roleplay language with a forced harm scenario. The second escalates from a threat to mutilate someone's body to a suggestion of repeatedly kicking them until they are near death. These examples show how violent content remains visible not only through explicit threats but through roleplay and fantasy violence.

\noindent
\textbf{\colorbox{black!20}{\#13:} Self-harm (ideation).} The following phrases indicate explicit expressions of self-harm or acute distress. Such language signals possible susceptibility and requires careful contextual interpretation. The prevalence of such content is harmful as it can normalize self-harm ideation in children.

\begin{tcolorbox}[colback=gray!7!white, colframe=black,
  boxrule=1pt,
  left=1mm,
  right=1mm,
  top=1mm,
  bottom=1mm,
  fontupper=\ttfamily\small
]
\footnotesize
\textbf{Self-Harm (Example Phrases)}\par\smallskip
\textbf{\texttt{Phrase 1:}}  \textcolor{red}{``i wish i can die''}\par
\textbf{\texttt{Phrase 2:}} \textcolor{red}{``HELP IM DYING''}\par
\textbf{\texttt{Phrase 3:}} \textcolor{red}{``I'm gonna drown myself''}\par
\end{tcolorbox}

\subsection{Sharing and Requesting PIIs}

Off-platform redirection and PII disclosure capture messages in which users attempt to move interactions beyond Roblox or share information that could help another user identify, locate, or contact them elsewhere. Next, we discuss our observations in this category.

\noindent
\textbf{\colorbox{black!20}{\#14:} Sharing or Requesting Off-platform Handles.}  These exchanges can be of two types: (1) a user sharing their own identity or (2) requesting others' identities in other platforms, such as TikTok, YouTube, and Snapchat. While not explicitly threatening or abusive, they remain pertinent because they create pathways for predators to track or continue communicating with children~\cite{zhao2019make}. For example, in 2024, a 15-year-old boy took his own life after shifting conversations to Discord, after severe manipulative coercion~\cite {Tan_2025}.

\begin{tcolorbox}[colback=gray!7!white, colframe=black,
  boxrule=1pt,
  left=1mm,
  right=1mm,
  top=1mm,
  bottom=1mm,
  fontupper=\ttfamily\small
]
\footnotesize
\textbf{Sharing Self-identity in Other Platforms}\par\smallskip

\texttt{\textbf{User1:} ``yo uu got a tt account''} $\rightarrow$
\texttt{\textbf{User1}: ``DO YALL WHATCH MY YOUTUBE''} $\rightarrow$
\texttt{\textbf{User2:} Do vou post codes \#\# \#\#} $\rightarrow$
\texttt{\textbf{User1:} Can v'all whatch it it's \textcolor{red}{[OFFPLATFORM\_HANDLE\_001]}} $\rightarrow$
\texttt{\textbf{User1:} \textcolor{red}{[OFFPLATFORM\_HANDLE\_001]}}\par
\end{tcolorbox}

\begin{tcolorbox}[colback=gray!7!white, colframe=black,
  boxrule=1pt,
  left=1mm,
  right=1mm,
  top=1mm,
  bottom=1mm,
  fontupper=\ttfamily\small
]
\footnotesize
\textbf{Requesting other's identity}\par\smallskip

\texttt{\textbf{User1:} ``m recording in tik''} $\rightarrow$
\texttt{\textbf{User2:} ``whats ur user''} $\rightarrow$
\texttt{\textbf{User1:} ``Its kinda long''} $\rightarrow$
\texttt{\textbf{User2:} ``ok tell me''} $\rightarrow$
\texttt{\textbf{User1:} \textcolor{red}{``[OFFPLATFORM\_HANDLE\_001]''}} $\rightarrow$
\texttt{\textbf{User2:} ``and post the tt after you tell and thx''}\par
\end{tcolorbox}


Interestingly, off-platform redirects appear through ordinary social conversation and not overtly suspicious language. In the first example above, a user asks whether others have a TikTok account, promotes a YouTube channel, and provides a recognizable account name. In the second, a user tells others they are recording for their channel, subsequently providing a period-linked username. While difficult to classify as harmful in isolated messages alone, their safety significance becomes clearer when read in context. Requests for usernames and references to outside platforms are extreme safety concern because the destination platform may have different moderation standards or detect coercive behavior in private messages. As a result, even seemingly ordinary requests to follow, post, or share usernames can increase children's exposure to interactions that occur outside the protections and visibility of Roblox's moderation environment.

\noindent
\textbf{\colorbox{black!20}{\#15:} Sharing Other's Sensitive Information.} Another not-so-prevalent but interesting observation is that users can use platforms like Roblox to share sensitive information on others -- third-party entities, who might not even be on Roblox. One such example is provided below, where one user shared another person's password.

\begin{tcolorbox}[colback=gray!7!white, colframe=black,
  boxrule=1pt,
  left=1mm,
  right=1mm,
  top=1mm,
  bottom=1mm,
  fontupper=\ttfamily\small
]
\footnotesize
\textbf{Possible Unauthorized Account Access}\par\smallskip

\textbf{\texttt{User1:}} \texttt{``he’s stupid so it must be an easv nass''} $\rightarrow$
\textbf{\texttt{User2:}} \texttt{``FINNALY DID IT''} $\rightarrow$
\textbf{\texttt{User3:}} \textcolor{red}{\texttt{\textbf{``BRO THE PASSWORD WAS [password]''}}} $\rightarrow$
\textbf{\texttt{User3:}} \texttt{``OK ITS DOWNLOADING''} $\rightarrow$
\textbf{\texttt{User3:}} \texttt{``HES GONNA NOTICE''} $\rightarrow$
\textbf{\texttt{User3:}} \texttt{``WELL IDC''} $\rightarrow$
\textbf{\texttt{User2:}} \texttt{``NOW''} $\rightarrow$
\textbf{\texttt{User3:}} \textcolor{red}{\texttt{``imma log into his acc''}}\par
\end{tcolorbox}

\subsection{Profanity}\label{sec:profanity}
Profanity captures messages containing explicit, implied, and vulgar swear words, at times appearing in partial or altered forms. 

\noindent
\textbf{\colorbox{black!20}{\#16:} Profanity in altered form.}
As shown below, most of the profanity examples use altered spellings, abbreviations, or substitute forms (e.g., ``\textit{f4},'' ``\textit{bi\$hs},'' and ``\textit{btx''}). 

\begin{tcolorbox}[colback=gray!7!white, colframe=black,
  boxrule=1pt,
  left=1mm,
  right=1mm,
  top=1mm,
  bottom=1mm,
  fontupper=\ttfamily\small
]
\footnotesize
\textbf{Examples of Profanity}

\vspace{0.5em}

\begin{tabular}{@{}p{0.49\linewidth} p{0.49\linewidth}@{}}
\textbf{P1.} \texttt{f4 off you dumb b4} &
\textbf{P2.} \texttt{allur bi\$hs} \\[0.1em]

\textbf{P3.} \texttt{Shithatfeelsgoodg} &
\textbf{P4.} \texttt{Dam imagine callin me a btc} \\[0.1em]

\textbf{P5.} \texttt{I look ugly as hell like this} &
\textbf{P6.} \texttt{ufelt so good f4 yeah ff} \\[0.1em]

\textbf{P7.} \texttt{she ndam gokng to wip Yourassi} &
\textbf{P8.} \texttt{HELL TO THE NO} \\
\end{tabular}

\end{tcolorbox}

Profanity was not limited to isolated swearing; it also appeared in exchanges involving casual emphasis, direct insults, sexual remarks, and aggressive escalation. Note that, here, we treat profanity as a category of unsafe policy violations that remained visible in the corpus. In RQ2 (\cref{sec:rq2 findings}), we return to these patterns from a different angle, examining how users adapted the language to bypass the moderation system.

\section{RQ2 (Evasion Techniques): Method} \label{sec:bypass_method}
This section describes the methods and results of Roblox users' attempts to evade moderation.
Since previously \textit{moderated} users are more likely to engage in evasive behaviors, we first identify moderated messages (i.e., masked by Roblox) and then examine subsequent behaviors of their senders. We manually review the full chat history of these users and perform thematic analysis to develop a taxonomy of evasion patterns, which we discuss next.


\noindent
\textbf{Moderated Chat Message Detection.} Roblox replaces potentially unsafe messages with sequences of hash (`\#') characters~\cite{moderation}; thus, we search for such sequences in the transcribed messages to identify moderated or sanctioned content. We observe that, since hash (`\#') is a special character, our OCR framework often recognizes the series of hashes as a series ``\verb|H|'' or ``\verb|4|''. To improve the precision of our moderated chat message detection, we score each span using rule-based features, such as length, repeated structure, character composition, and penalties for forms that resemble common words or likely false positives (e.g., ``\verb|AHHHH|'' or ``\verb|HAHAHA|''). 
%
We exclude masked spans of length one or two, as these spans are disproportionately difficult to distinguish from low-information moderation events. Instead, we prioritize masked spans of length three or greater, as they likely enhance support for sequence-level analysis of bypass behavior. Table~\ref{tab:game_hash_profile} summarizes the distribution of masked content across games with observed user count. Figure~\ref{fig:hash-length-distribution} in~\cref{hash:length} shows the distribution of masked content of different lengths.

\begin{figure}[h]
    \centering
    \includegraphics[width=\columnwidth]{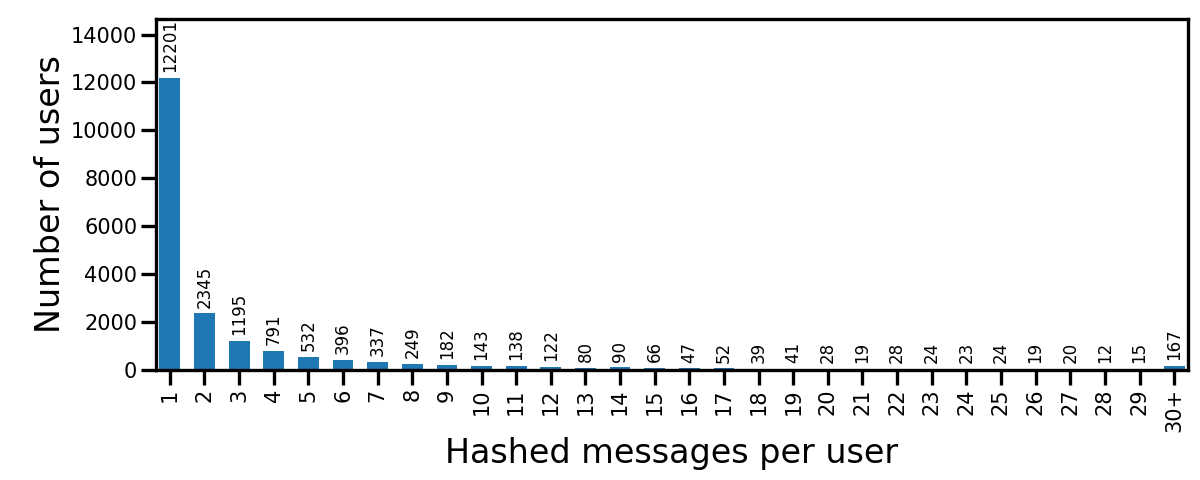}
    \caption{Moderated/hashed message count per user.}
    \label{fig:rank_frequency_hashed_users}
\end{figure}

\noindent
\textbf{Thematic Analysis to Understand Evasion Patterns.}\label{sec:manual_review}
Next, we manually review masked content and surrounding messages based on \textit{users} to surface evasive behaviors. Due to the large volume of masked content, we prioritize the ones most likely to support meaningful qualitative interpretation. \Cref{fig:rank_frequency_hashed_users} shows how the rank distribution of users was highly skewed, with a small subset of users accounting for the bulk of moderated messages. Therefore, to reduce the overrepresentation of a small number of highly active users or a single game environment, we select review targets in a stratified manner across games and user-level frequency groups. We use a cutoff at users with at least seven masked messages to focus our effort on high-information-density users.
%
%
When we assign users to frequency groups, we do so based on the proportion of their hashed messages that met our confidence criteria. Specifically, we compute a \underline{\textit{frequency ratio}} by dividing the number of hashed messages by the total number of messages. Users with a ratio of above 0.90 are assigned to the high-frequency group. Users with ratios between 0.50 and 0.90 are assigned to the medium-frequency group, and users with ratios of 0.50 or below are assigned to the low-frequency group. Note that Roblox's moderation is often partial, which means only part of the message is masked; thus, it is still valuable to analyze users with a high frequency ratio. Table \ref{tab:sample_candidate_users} shows some examples of users selected for manual review.


\begin{table}[h]
\centering
\caption{Representative candidate users selected for manual review.}
\label{tab:sample_candidate_users}
\small
\setlength{\tabcolsep}{4pt}
\begin{tabular}{llrrrr}
\toprule
Game & User & Freq. & Masked & Freq.\\
     &      & Level & Msgs.  & Ratio\\
\midrule
AdoptMe    & U15127 & High   & 11 & 1.00\\
AdoptMe    & U09489 & Medium & 18 & 0.89\\
BerryAve   & U03241 & High   & 12 & 0.92\\
BerryAve   & U32259 & Low    & 14 & 0.43\\
Brookhaven & U75564 & High   & 11 & 0.91\\
Brookhaven & U62192 & Medium & 18 & 0.83\\
RoyaleHigh & U86959 & High   & 16 & 1.00\\
RoyaleHigh & U11452 & Medium & 15 & 0.73\\
\bottomrule
\end{tabular}
\end{table}



The selected masked content and surrounding messages were then manually annotated by following an iterative thematic analysis process.
Following prior taxonomy work~\cite{Mpofu_16094069251348542, educsci13020098, Chen_Anandayuvaraj_Davis_Rahaman_2024}, we used saturation as a stopping condition. We consider saturation has been reached when all three of the following conditions are satisfied: 
\begin{itemize}
    \item Including a newly reviewed candidate does not expand the set of unique evasion techniques or alter the defining properties of an existing theme. Saturation is the coupling of diminishing returns and information redundancy, and not evidence that all possible techniques have been exhausted.
    \item The focal candidate must contain sufficient interpretable context to support comparison with previously reviewed cases. This criterion prevents saturation from being declared on the basis of sparse or uninformative cases.
    \item Both conditions must hold for three consecutive candidate windows. We add this condition to provide additional safeguard against premature saturation caused by incidental redundancy in the review sequence.
\end{itemize}

In total, we analyzed 12,612 messages from 94 users until thematic saturation was reached.

\section{RQ2 (Evasion Techniques): Findings} \label{sec:rq2 findings}

This section summarizes the recurring evasion techniques users employed to adapt their language after content was filtered. Similar to \cref{subsubsec:rq2haconclusions}, relevant contents supporting the narratives are highlighted in \textcolor{red}{red}, and anonymized contents are highlighted in \textcolor{red}{[text]}.



\subsection{Multi-Line Fragmentation}
\textit{Multi-line fragmentation} captures cases where users attempted to preserve meaning by distributing fragments of filtered or partially filtered messages across adjacent lines (turns). Rather than replacing a blocked word with a single alternate form, users repeated, continued, or repaired the filtered message over several consecutive messages. This pattern appeared most frequently in the reviewed saturation set due to the sequential structure of the chat stream, allowing for active response to filtered feedback.

\noindent
\textbf{\colorbox{red!20}{\#1:} Bypassing for whole sequences.}
In some cases, the original message was entirely filtered, to only later reconstruct the intended meaning in subsequent turns. The following example shows a heavily masked sequence, accompanied by partial fragments, materializing the self-destructive language. 
Multi-line fragmentation effectively bypasses sentence-level moderation, as users can distribute meaning across multiple lines, revealing intent even some individual lines are partially masked.

\begin{tcolorbox}[colback=gray!7!white, colframe=black,
  boxrule=1pt,
  left=1mm,
  right=1mm,
  top=1mm,
  bottom=1mm,
  fontupper=\ttfamily\small
]
\footnotesize
\textbf{(b) Multi-line retry when whole sentence was masked}\\
\texttt{\textbf{User:} \#\#\#\#\#\#\#\#\#\#\#\#\#\#\#\#\#\#} $\rightarrow$
\texttt{\textbf{User:} ``I just want to''} $\rightarrow$
\texttt{\textbf{User:} \#\#\#} $\rightarrow$
\texttt{\textbf{User:} \#\#\#} $\rightarrow$
\texttt{\textbf{User:} ``die''}
\end{tcolorbox}

\noindent
\textbf{\colorbox{red!20}{\#2:} Bypassing for partial sequences.}
In other cases, only part of the original message was filtered. In this example, an insult sequence begins with a partially masked accusation, then continues across ensuing texts: \textit{CALLED ME}, \textit{A}, \#\#\#\#\#, and finally \texttt{``BTC''}. The surrounding sequence narrows the interpretation and makes the intended profanity legible through context. This illustrates how users can use feedback over neighboring turns for bypassing.

\begin{tcolorbox}[colback=gray!7!white, colframe=black,
  boxrule=1pt,
  left=1mm,
  right=1mm,
  top=1mm,
  bottom=1mm,
  fontupper=\ttfamily\small
]
\footnotesize
\textbf{(c) Multi-line retry when partial sentence was redacted}\\
\texttt{\textbf{User:} ``NAWW SOMEONE \#\#\#\#\#\# \#\# \# \#\#\#\#\#''} $\rightarrow$
\texttt{\textbf{User:} ``CALLED ME''} $\rightarrow$
\texttt{\textbf{User:} ``A''} $\rightarrow$
\texttt{\textbf{User:} \#\#\#\#\#} $\rightarrow$
\texttt{\textbf{User:} ``BTC''}\par
\end{tcolorbox}

\noindent
\textbf{\colorbox{red!20}{\#3:} Bypassing for a sequence.}
\textit{Multi-line} fragmentation also appeared when users split a filtered word in multiple turns. In the following example, the exchange begins with an ordinary age-related question, then escalates to a request for contact information. The requester uses a shorthand symbol reference, ``\textit{\#},'' then a masked message indicating the more direct version may have triggered moderation. Instead of stopping, the user repairs the request in the next two turns.

\begin{tcolorbox}[colback=gray!7!white, colframe=black,
  boxrule=1pt,
  left=1mm,
  right=1mm,
  top=1mm,
  bottom=1mm,
  fontupper=\ttfamily\small
]
\footnotesize
\textbf{(a) Multi-line retry}\\
\texttt{\textbf{User1:} ``that’s cool! What grade u in now?''} $\rightarrow$
\texttt{\textbf{User2:} \textcolor{red}{[6th]}} $\rightarrow$
\texttt{\textbf{User2:} ``I’m young''} $\rightarrow$
\texttt{\textbf{User1:} ``u have a \#''} $\rightarrow$
\texttt{\textbf{User1:} ``A what''} $\rightarrow$
\texttt{\textbf{User1:} \#\#\#\#\#\#} $\rightarrow$
\texttt{\textbf{User1:} ``numbe''} $\rightarrow$
\texttt{\textbf{User1:} ``r''}
\end{tcolorbox}

\subsection{Lexical Retry}
Lexical retry portrays cases where users responded to moderation by restating blocked content in a modified form. Instead of reiterating the same message in later turns, users changed the wording, spelling, or representation of the original content while still preserving the underlying intent. This pattern distinguishes reformulation from repetition by illustrating that a message may change on the surface, but the communicative goal remains the same.

\noindent
\textbf{\colorbox{red!20}{\#4:} Bypassing through word reformat.}
In some cases, users reformulated filtered information by revising its representation. In the following example, a user initially sends a numeric sequence, parts of which were masked by the third line, then continued spelling out numbers as words. The retry, therefore, did not introduce new content but changed the surface form of sensitive information after moderation. This bypass technique is useful for communicating sensitive information, such as phone numbers.

\begin{tcolorbox}[colback=gray!7!white, colframe=black,
  boxrule=1pt,
  left=1mm,
  right=1mm,
  top=1mm,
  bottom=1mm,
  fontupper=\ttfamily\small
]
\footnotesize
\textbf{(a) Lexical Retry For Sending Numbers}\\
\texttt{\textbf{User:} ``A code''} $\rightarrow$
\texttt{\textbf{User:} ``Its''} $\rightarrow$
\texttt{\textbf{User:} }\textcolor{red}{``[234]''} $\rightarrow$
\texttt{\textbf{User:} }\textcolor{red}{``[567]''} $\rightarrow$
\texttt{\textbf{User:} \#\#\#} $\rightarrow$
\texttt{\textbf{User:} \#\#\#\#} $\rightarrow$
\texttt{\textbf{User:} }\textcolor{red}{``[Four]''} $\rightarrow$
\texttt{\textbf{User:} }\textcolor{red}{``[Three]''} $\rightarrow$
\texttt{\textbf{User:} }\textcolor{red}{``[Two]''}
\end{tcolorbox}

\noindent
\textbf{\colorbox{red!20}{\#5:} Bypassing through sentence revision.}
In other cases, users responded to filtering by revising the sentence rather than splitting the content over subsequent turns. In this example, the user attempts to relay their name, but the original term is partially filtered. The user then reformulates the sentence, preserving the same disclosure while changing the ordering.

\begin{tcolorbox}[colback=gray!7!white, colframe=black,
  boxrule=1pt,
  left=1mm,
  right=1mm,
  top=1mm,
  bottom=1mm,
  fontupper=\ttfamily\small
]
\footnotesize
\textbf{(b) Lexical Retry For Other Sensitive Info}\\
\texttt{\textbf{User:} ``My name is also \#\#\#\#\#''} $\rightarrow$
\texttt{\textbf{User:} \textcolor{red}{``[Sarah]} is mv name too''}
\end{tcolorbox}

\subsection{Altered Spelling}
Altered spelling captures cases where users truncate or substitute the written form of a filtered term. These alterations include phonetic spellings, added or removed letters, and spacing changes. Unlike lexical retry, which may revise the sentence, altered spelling operates primarily at the word level. The user keeps the same underlying term but modifies how it appears on the screen.

\noindent
\textbf{\colorbox{red!20}{\#6:} Bypassing for profane language.}
In some cases, users used adapted spelling to preserve profane language after filtering. In the following example, a filtered phrase was reissued in a version that added a period after the determiner, then altered the spelling of the profane term. Rather than abandoning the blocked expression, the added punctuation disrupted the original phrase structure, while the nonstandard spelling changed the word form itself.

\begin{tcolorbox}[colback=gray!7!white, colframe=black,
  boxrule=1pt,
  left=1mm,
  right=1mm,
  top=1mm,
  bottom=1mm,
  fontupper=\ttfamily\small
]
\footnotesize
\textbf{(a) Altered Spelling}\\
\texttt{\textbf{User:} ``I \#\#\#\# \#\#\# ake \#\#\#\# in your lab''} $\rightarrow$
\texttt{\textbf{User:} ``I need to take a. \textcolor{red}{shi} in your lab''}
\end{tcolorbox}

\noindent
\textbf{\colorbox{red!20}{\#7:} Bypassing for sexually explicit language.}
Users also altered spelling to sexually explicit terms. In this example, the user begins with a nonstandard spelling of the term, encounters filtering, then reissues the term in a phonetically distorted form. Although the variation evolves across turns, the user modified the word's surface to avoid the exact spelling that triggered masking.

\begin{tcolorbox}[colback=gray!7!white, colframe=black,
  boxrule=1pt,
  left=1mm,
  right=1mm,
  top=1mm,
  bottom=1mm,
  fontupper=\ttfamily\small
]
\footnotesize
\textbf{(b) Altered Spelling}\\
\texttt{\textbf{User:} ``eat ur posay ma'am''} $\rightarrow$
\texttt{\textbf{User:} \#\#\#\#\#\#} $\rightarrow$
\texttt{\textbf{User:} ``\textcolor{red}{pOHSSAY}''}
\end{tcolorbox}

\subsection{Code Word}
Code word cases involve substitute forms, abbreviations, or shortened terms. Unlike altered spelling or leet speak, code words do not preserve the original visual form of the word. Instead, they rely on context, sequence, and shared interpretation. This category also connects with the escaped category analysis in \cref{sec:profanity}, where forms such as \textit{f4} and \textit{btx} illustrate how numbered substitutes and married keys (``\textit{x}'' and ``\textit{c}''), allowed profane language to remain in chat. 

\noindent
\textbf{\colorbox{red!20}{\#8:} Bypassing for profanity.}
All the instances of code words that we observed are used to express expletive insults. In the following examples, users first produced a partially filtered message, then replaced it with a shortened substitute such as \textit{b} or \textit{BTC}. The preceding messages prepare the interpretation, and the substitute term supplies the recoverable meaning.

\begin{tcolorbox}[colback=gray!7!white, colframe=black,
  boxrule=1pt,
  left=1mm,
  right=1mm,
  top=1mm,
  bottom=1mm,
  fontupper=\ttfamily\small
]
\footnotesize
\textbf{(a) Code Word For Profanity}\\
\texttt{\textbf{User:} ``u ugly \#\#\#\#''} $\rightarrow$
\texttt{\textbf{User:} ``mind ur business''} $\rightarrow$
\texttt{\textbf{User:} ``u \textcolor{red}{b}''} $\rightarrow$
\texttt{\textbf{User:} ``yea run away \textcolor{red}{b}''}
\end{tcolorbox}

\begin{tcolorbox}[colback=gray!7!white, colframe=black,
  boxrule=1pt,
  left=1mm,
  right=1mm,
  top=1mm,
  bottom=1mm,
  fontupper=\ttfamily\small
]
\footnotesize
\textbf{(b) Another Example of Code Word for Profanity} \\
\texttt{\textbf{User:} ``NAWW SOMEONE \#\#\#\#\#\# \#\# \# \#\#\#\#\#''} $\rightarrow$
\texttt{\textbf{User:} ``CALLED ME''} $\rightarrow$
\texttt{\textbf{User:} ``A''} $\rightarrow$
\texttt{\textbf{User:} \#\#\#\#\#} $\rightarrow$
\texttt{\textbf{User:} ``\textcolor{red}{BTC}''}
\end{tcolorbox}


In the final example below, the user moves through several hostile turns using terminology as surrogates for profane language in an escalating exchange. The surrounding phrases make their function clear, for example, ``\textit{f4}'' appears inside the familiar phrase ``\textit{get the [f-word] out}.'' This makes substitution more direct than a vague code word, as the sentence frame strongly identifies what kind of word belongs there. 

\begin{tcolorbox}[colback=gray!7!white, colframe=black,
  boxrule=1pt,
  left=1mm,
  right=1mm,
  top=1mm,
  bottom=1mm,
  fontupper=\ttfamily\small
]
\footnotesize
\textbf{Direct textual substitution}\\
\texttt{\textbf{User:} ``Your not involved btc''} $\rightarrow$
\texttt{\textbf{User:} ``Get the hell out''} $\rightarrow$
\texttt{\textbf{User:} ``Before shii get ugly''} $\rightarrow$
\texttt{\textbf{User:} ``Get the \textcolor{red}{f4} out''}
\end{tcolorbox}

\subsection{Leet Speak}
Leet speak replaces standard letters with visually similar numbers, special characters, or other orthographic variations~\cite{stano2023linguistic}. Leet exists on a spectrum, where basic forms are often readable to the human eye (e.g. \textit{h4cker} or \textit{n00b}), and harder forms use denser symbol substitution making them difficult to detect. This makes leet useful given that it exploits the gap between how people read a language and how automated systems process text as literal string matching.

\noindent
\textbf{\colorbox{red!20}{\#9:} Bypassing with symbolic substitution.}
In some cases, leet speak appeared as a character-level strategy for modifying filtered words. In the following example, a user replaces the letter ``\textit{a}'' with ``\textit{@}'' resembling the letter shape in some stylized writing forms.

\begin{tcolorbox}[colback=gray!7!white, colframe=black,
  boxrule=1pt,
  left=1mm,
  right=1mm,
  top=1mm,
  bottom=1mm,
  fontupper=\ttfamily\small
]
\footnotesize
\textbf{(a) Leet Speak}\\
\texttt{\textbf{User:} \#\#\#\#\#\#\#\#\#\#\#\#\#\#\#\#\#\#} $\rightarrow$
\texttt{\textbf{User:} ``what's your''} $\rightarrow$
\texttt{\textbf{User:} ``\textcolor{red}{n@me}''} $\rightarrow$
\end{tcolorbox}

\noindent
\textbf{\colorbox{red!20}{\#10:} Bypassing with numeric substitution.}
In the next example, leet speak occured through numerical substitutions. Words containing leet digits can activate the intended word base with little processing, explaining why forms similar to \textit{CHAT5} and \textit{d8ing} regularize into a letter or sound based on its position~\cite{Perea_2008}. In \textit{CHAT5}, the ``\textit{5}'' functions as an ``\textit{S}'' due to its visual resemblence. The ``\textit{8}'' in ``\textit{d8ting}'' is a phonetic compression with ``\textit{eight}'' approximating the sound sequence in ``\textit{ate}.''

\begin{tcolorbox}[colback=gray!7!white, colframe=black,
  boxrule=1pt,
  left=1mm,
  right=1mm,
  top=1mm,
  bottom=1mm,
  fontupper=\ttfamily\small
]
\footnotesize
\textbf{(b) Leet Speak}\\
\texttt{\textbf{User:}  ``I \textcolor{red}{SCRE2ENSHOTTED CHAT5''}} $\rightarrow$
\texttt{\textbf{User:}  ``Y’all are \textcolor{red}{d8ing}''}
\end{tcolorbox}

\subsection{Probing}
Probing captures cases where users test multiple variants of a blocked term or phrase. Unlike lexical retry, where users redesign the filtered message in format and sentence revision, these examples involve an explicit trial-and-error process attempting alternate forms such as abbreviations, descriptions, or partial spellings. Using this technique, users explore information about what the moderation system will allow or disallow.

\noindent
\textbf{\colorbox{red!20}{\#11:} Bypassing for off-platforms.}
In some probing cases, users attempt to reference an off-platform application, only to have the moderation system completely masked the name. In the following example, after the initial question, ``\textit{do yall have \#\#\#\#\#\#\#},'' the user attempts to compress the name through abbreviation and comments on the outcome. This comment is important because it shows awareness of the filter as an active constraint in the conversation. The user does not merely repeat the blocked work, but adjusts their next trail to describe it as starting with ``\textit{disco}'' and ending in ``\textit{rd}'' for the completed spelling ``\textit{discord}''. Each step supplies new information while avoiding the exact form that previously triggered moderation. This type of probing offers a significant challenge to the robustness of the moderation system.

\begin{tcolorbox}[colback=gray!7!white, colframe=black,
  boxrule=1pt,
  left=1mm,
  right=1mm,
  top=1mm,
  bottom=1mm,
  fontupper=\ttfamily\small
]
\footnotesize
\textbf{(a) Probing for Moving Off-platform}\\
\texttt{\textbf{User:} ``do yall have \#\#\#\#\#\#\#''} $\rightarrow$
\texttt{\textbf{User:} ``DSC''} $\rightarrow$
\texttt{\textbf{User:} ``WHY DOES \#\#\#\#\#\#\# TAGS''} $\rightarrow$
\texttt{\textbf{User:} ``the app that starts with \textcolor{red}{disco}''} $\rightarrow$
\texttt{\textbf{User:} ``ends with''} $\rightarrow$
\texttt{\textbf{User:} ``\textcolor{red}{rd}''}
\end{tcolorbox}

\noindent
\textbf{\colorbox{red!20}{\#12:} Bypassing for geographical information.}
In other probing cases, users test whether geographical information could pass through filtering. Unlike ordinary location sharing, these examples show users actively negotiating what could be said, where it could be placed, and whether the system would continue to mask it. In this example, the exchange begins with a question concerning a specific city, then a partially masked phrase before \textit{Houston} (anonymized), suggesting that the moderation system responds consistently to reference or surrounding wording. The unevenness encourages additional experimentation with the user moving to question the other's current time. This probing behavior creates a privacy risk because location disclosure can support off-platform contact, identification, or real-world harm.

\begin{tcolorbox}[colback=gray!7!white, colframe=black,
  boxrule=1pt,
  left=1mm,
  right=1mm,
  top=1mm,
  bottom=1mm,
  fontupper=\ttfamily\small
]
\footnotesize
\textbf{(b) Probing for Sharing Location Information}\\
\texttt{\textbf{User1:} ``you live in \textcolor{red}{[El Paso]} right''} $\rightarrow$
\texttt{\textbf{User1:} ``do you live in \textcolor{red}{[El Paso]}?''} $\rightarrow$
\texttt{\textbf{User2:} ``yea''} ...
\texttt{\textbf{User1:} ``bc the time is diffrent in the diff cities''} $\rightarrow$
\texttt{\textbf{User3:} \# \#\#\#\# \#\# \#\#\#\#\#\#\# \textcolor{red}{[Houston]}} $\rightarrow$
\texttt{\textbf{User3:} ``what time is it right now for u''}
\end{tcolorbox}

%% file: Main_document/discussion.tex
\section{Discussion}
\label{sec:discussion} 
In this paper, we analyzed a large-scale, real-world chat corpus from popular games on Roblox. Our analysis identified a wide range of harmful content that evaded Roblox's moderation system. Categorization of these messages shows that grooming dominates harmful behaviors, followed by bullying and threats to violence (\cref{tab:unsafe-combined}). This is troublesome, but perhaps unsurprising, given that the popularity of these games among minors may attract predators.

Our analysis further detected evasive tactics that users adopted to sneak in harmful content. These include well-known techniques like leet speak and algospeak, which are popular on other online platforms, as well as techniques likely invented specifically for the conversational settings Roblox games offer, such as splitting the content into multiple turns and trying out variants of a moderated message. 

\subsection{Implications of our findings} 
Our findings emphasize the limitations of message-level moderation in capturing conversational and context-dependent harm. This highlights the need for context-aware moderation that is capable of filtering multi-turn conversations. Note that during our pre-filtering step, the best model achieved a recall of 72.4\% (\cref{tab:llm_validation}), which means it missed one in every four messages; thus, harmful interactions are more prevalent than what this result shows. In addition to the observed evasion strategies, we would like to highlight one other observation, as \cref{fig:rank_frequency_hashed_users} shows, a small number of users are responsible for the bulk of the harmful content, and they are persistent in bypassing the moderation. This implies that moderation decisions are largely made at the message level without incorporating user-level behavioral context, allowing persistent offenders to repeatedly bypass safeguards and continue engaging in harmful interactions.

\subsection{Recommendations}\label{sec:recommendation}
In light of our findings, we propose the following recommendations:

\noindent
\textbf{\colorbox{green!20}{Recommendation \#1:} mix deterministic and probabilistic approaches.} Many evasive techniques used spelling variants of the harmful word or phrases, such as mixing characters and symbols or intentionally misspelling. Deterministic approaches---like regular expressions and approximate string matching coupled with regularly updated datasets of popular slangs and codewords---might detect them better than probabilistic language models, and within a shorter time and with less computing resources. Language models can then focus on more serious cases, such as grooming.   

\noindent
\textbf{\colorbox{green!20}{Recommendation \#2:} judge based on `conversation' rather than individual messages.} As our data shows, often users split a single message into multiple turns to bypass moderation. We also found that certain messages can be determined as abusive only when considered within the context or implicit cues. Thus, instead of judging messages in isolation, maintaining a window of conversational messages is required to improve the detection accuracy. This is easier said than done: focusing on the continuous conversation stream will dramatically increase computing demand, and can be unsustainable at the million-user scale. Rather, continuous monitoring may focus on a few high-risk users, as explained next.

\noindent
\textbf{\colorbox{green!20}{Recommendation \#3:} longitudinal and cross-game examination of users.} We found that a small number of users post bulk of the offensive messages, and repeatedly get filtered by the existing moderation system. Roblox could deploy a more sophisticated moderation strategy (e.g., a better but slower language model that supports longer context, possibly assisted by human moderators) reserved for users whose messages had already been moderated beyond a certain threshold. User-level information could be shared across servers or games, so that someone who was blocked from one game would face stricter moderation in another. Indeed, Roblox game creators have demanded such platform-level moderation assistance~\cite{kou2025system}, noting the challenges in coordinating and exchanging information within the current setup.

\noindent
\textbf{\colorbox{green!20}{Recommendation \#4:} improve the reporting mechanism and build trust among the users.} The mechanism to report misbehaving users is the fallback mechanism when moderation fails. Roblox should strengthen and promote its reporting mechanism to create awareness of it~\cite{kou2025system}. We also suspect that there is a lack of trust in the system, i.e., users may not believe that by reporting, they will achieve anything. This is justifiable as they see abusive messages getting filtered (indicating that the system can detect them), but the abuser is still able to send more messages. 
Thus, we also recommend that platforms build trust by showing that reporting users help make the platform safer.

\subsection{Study Limitations}


Our team could only collect publicly available data from public-facing game servers; thus, our results are not generalizable to private servers. Additionally, because Roblox attracts underage players, many messages reflect early writing skills that include grammatical errors, misspellings, acronyms, and jargon. While we believe we can understand most discussions, some generational colloquialisms may not be recognized. Finally, video recordings rotated servers periodically to prevent freezing, game updates, and UI changes, which may introduce minor content disruption between sessions. After transcription, the textual data retained some noise, particularly in cases involving non-English languages or emoticons. 

We also note that due to ethical reasons, we used \textit{gpt-oss-120B} with few-shot prompting with an F1 score of 44.13\%, with 31.39\% precision and 71.43\% recall. The impact of low precision on our overall result is minimal, as it only increased our effort during manual thematic analysis. However, 71.43\% recall indicates a fair amount of unsafe messages escaped our pre-filtration process. Similarly, our saturation-based analysis might also miss some rare patterns. Thus, our findings should be interpreted as a lower-bound estimate on the true prevalence of unsafe interactions.

%% file: Main_document/relatedwork.tex
\section{Related Work}
\label{sec:relwork}

\noindent
\textbf{Content Moderation in Online Platforms.}
 Online content moderation began largely as human volunteer-driven processes~\cite{10.1145/3290605.3300390, matias2019civic, kayany1998contexts, 1241285}, where rules were often informal and enforced through social norms or ad hoc judgments ~\cite{de2021peer, annamoradnejad2022requirements}. As platforms grew, moderation systems evolved by establishing centralized control ~\cite{yang2021tar, annamoradnejad2022requirements, seering2019moderator}. Social media and gaming platforms shifted towards automated moderation~\cite{gillespie2020content, makhijani2021quest} with formal guidelines and reporting systems~\cite{cai2024content, 10190527,  10.1145/3406865.3418312}. These platform-automated approaches spanned keyword filtering, machine learning classifiers~\cite{DBLP:conf/icmla/AhmedHS23, an2025toward}, and AI-based systems~\cite{yang2025realfactbench, Kumar_2024, Thomas_2025} designed to identify harmful content~\cite{qureshi2025explainable, goyal2025momoe}. %

 \noindent
\textbf{Automated Approaches to Child Safety.}
Comparatively little research examines how violations unfold within MMO environments. Existing work that focuses on age-inappropriate content~\cite{livingstone2012children, DBLP:conf/icmla/AhmedHS23, flynn2024child, hong2024impact, eltaher2025protecting, radesky2024algorithmic}, cyberbullying~\cite{hinduja2013social, mccrae2017social, aliyeva2025deep}, or predatory behavior~\cite{engelmann2025developing, whittle2013review, bihani2025fuzzy, finkelhor2022prevalence}, fails to develop intervention methods that shield children from online abuse. Legal frameworks such as COPPA~\cite{coppareg} mandate age filters and parental consent, yet studies report children often circumvent age-verification tools~\cite{pasquale2020digital} and parental controls are inconsistently applied ~\cite{robloxparental, instagramparental}. Recent advances in machine learning, LLMs, and computer vision offer new ways to identify harmful content at scale ~\cite{Hasan_Athrey_Khalid_Xie_Younessian_Braskich_2024, 10680313, Dendi_2025}. Yet automated classifiers trained on toxicity benchmarks struggle with nuanced aggression, age-inappropriate content, and manipulation~\cite{computation13080196, 10190527, Garcia_Carvalho_2025, yousaf2022deep}. Computer vision demonstrates that OCR can reliably extract text embedded within images, and then classify messages using deep learning techniques~\cite{Briskilal_2024, cmes.2025.061653, 11323612}. Yet these pipelines still struggle with text orientations, lighting, and background noise that require continuous human-in-the-loop review ~\cite{11168676, 11140757}. 

\noindent
\textbf{Positioning Our Work.}
Our work fills several gaps in the existing literature. Prior works mainly focused on online networking platforms (e.g.,~\cite{gatta2023interconnected, goldstein2023understanding}) that provide a different interaction environment than MMO platforms. The latter is characterized by interactions among a small number of users in a private or semi-private environment hidden from public view. Furthermore, MMO platforms, and in particular Roblox, have a disproportionately large number of underage users. Thus, the distribution of different harmful content and their severity are expected to vary; in particular, MMO platforms may be abused for grooming and sexual or mental abuse of minors. However, no prior work has evaluated how effectively current moderation techniques can prevent such abuses. 
Additionally, prior audits examined moderating actions~\cite{juneja2023, Hong_corona, goldstein2023, Trujillo_2025} but did not study how users adapt to bypass moderation. Our study addresses this gap by identifying tactics people use to evade detection while sending harmful messages. Finally, research on chat or messaging platforms have predominantly used crowdsourced data, which is susceptible to selection bias~\cite{gonccalves2026potential, razi2023sliding}. To our knowledge, this is the first study using chat messages collected in the wild; the findings are thus expected to better represent the reality.





\section{Conclusion}
We conduct the first large-scale analysis of chat messages on Roblox. This study surfaces the limitations of Roblox's current chat moderation pipeline, and importantly, suggests that any improvement in moderation will be followed by users inventing new tactics to evade it. We conclude by strongly recommending a user-centered (rather than individual message-centered), layered moderation pipeline, where behaviors of the same user across sessions, or even across games, need to be evaluated collectively, with an increasingly sophisticated mechanism after each flagged instance.   

\section{Acknowledgements}
This paper was edited for grammar using Grammarly~\cite{grammarly}. Sample videos and images provided in this paper and in the artifacts were generated through ChatGPT and Gemini Pro 3.1~\cite{chatgpt, chatbot_app} for anonymization purposes. This project benefited from many contributors who deserve recognition. We would like to extend our gratitude to the current and former members of SPRLab at the University of Arizona, including Jesse Chen, Rubin Yang, Muhammad Bilal, Talha Abrar, Md Moyeen Uddin, Saiful Islam Salim, and Xin Li, for helpful discussions and feedback on the manuscript. Special thanks to Saiful for suggesting a computer vision-based approach for data collection and to Saiful, Jesse, and Rubin for helping Priya during the initial stages of data collection and analysis. Finally, we thank Mihai Surdeanu for letting us use his lab's computing resources for this project.

%% file: Main_document/conclusion.tex


%% file: Main_document/appendix.tex



\section{Ethical Considerations}\label{ethics}

This study involves collecting chat messages involving minors from third-party servers. Even if these servers are public, i.e., any registered users can access them, adhering to the highest ethical standards is paramount for ensuring the protection, privacy, and well-being of the individuals, especially minors, represented in the dataset~\cite{DBLP:conf/icwsm/FieslerBK20}. Thus, we worked with our Institutional Review Board (IRB) to design a study protocol that responsibly manages all the ethical and legal requirements. The protocol underwent a full board review before it was approved, with the condition that any changes be reported within 6 months, which we complied with. One significant concern with the protocol was the lack of consent from game server owners or the individuals involved in the chat. Previous studies have shown that revealing the nature of observational studies like ours can significantly alter users' actions~\cite{wickstrom2000hawthorne}, undermining the purpose of the study. Also, prior work shows that obtaining consent in chat environments like ours is impractical~\cite{hudson2004go}, as either researchers were banned from the chatrooms or only a small number of people consented, rendering the study statistically unrepresentative. An attempt to obtain consent after data collection would be impractical, given that 105,214 users were involved and could risk re-identifying individuals. 

We argue that despite the absence of explicit consent, the benefits of this study outweigh the risks, given the careful and proactive measures we take to mitigate potential harms. Firstly, we designed a \textit{non-intrusive} method for data collection to comply with the Terms of Use of Roblox, including the ``License to and Restrictions of Services'' provisions~\cite{robloxRobloxTerms}. In accordance with these terms, the study did not involve any reverse engineering, disassembly, or modification of Roblox's technology, nor did it access any restricted servers. Secondly, 
data involving minors introduces risks such as unintended re-identification, exposure of sensitive content~\cite{proferes2021studying}, and potential misuse if harmful interaction patterns are reproduced without safeguards. Secondly, to mitigate these concerns, we apply deanonymization as soon as we transcribe the chat messages for downstream analysis and store the raw data on a password-protected, air-gapped hard drive. Thirdly, to prevent data exposure to third-party LLM services, we only used open-source local LLMs in this study. Fourthly, we selectively redacted or minimally quoted the examples to preserve analytical value while preventing traceability in the paper. We do not release raw datasets and instead report only aggregated findings. Finally, we are in the process of responsibly disclosing our findings to Roblox, including examples of moderation bypasses, recurring evasion patterns, and harmful interaction patterns identified in our analysis, in the hope of helping them design a better moderation system that will positively impact all users on their platform.

\section{Appendix}

\subsection{Anonymization of Usernames}\label{anonymization:dataset}
To anonymize the usernames found in the recording transcripts, we designed a script to first identify common name formats and OCR variants (e.g., brackets, colons, pipes, and punctuation). The role tags, such as ``\texttt{[VIP]}'' and ``\texttt{[Team]},'' were removed to avoid being mistaken for usernames. Once a possible username was found, the script normalized the text by lowercasing it, removing accents/extra punctuation, collapsing spaces, and keeping only letters, numbers, and underscores. Next, to avoid treating OCR errors as different monikers, the script compared newly detected usernames to previously identified ones. If a new candidate was >90\% similar to a known username, the script reused the known username rather than creating a separate identity. This threshold helped account for small OCR distortions, such as missing characters, extra punctuation, or visually similar characters, while reducing the chance of merging unrelated users. After parsing and normalization, each unique username was replaced with a pseudonym such as ``\texttt{user\_00001},'' and ``\texttt{user\_00002}.'' These pseudonyms were assigned globally across all files, so the same detected user kept the same anonymized label wherever they appeared. Lines that were identified as server messages were labeled ``\texttt{server},'' and lines where the speaker could not be confidently identified were labeled ``\texttt{unknown}.'' The anonymized output kept only the pseudonymous speaker label and message text. A separate username mapping file was created for internal auditing, ensuring analysis focused exclusively on de-identified data. This allowed us to preserve conversational structure and repeated speaker patterns without exposing original Roblox usernames.



\section{Additional Steps for Data Processing \& Analysis}

\subsection{Background-Suppression}\label{background:removal}

We apply background suppression using an empirically adjusted RGB threshold combination. To obtain an optimal threshold for the background suppression, we generated RGB masks thresholds using values drawn from the set \{50, 100, 150, 200\}, yielding 64 candidate RGB threshold combinations (e.g., ${(50, 50, 50), (50, 50, 100)}$ ${\dots (200, 200, 200)}$). We applied these RGB thresholds on the ground-truth set to generate respective background-suppressed images for all 64 variants. We then generate text using the OCR for the ground-truth set and compare the OCR-generated text against the text from the ground-truth set for all variants. On average, the best performing variant (for all games) achieves an accuracy of approximately $90\%$, with $80\%$  lines successfully matched (both numbers averaged for all games). Accordingly, the best-performing RGB variant is selected on a per-game basis and applied across the corresponding corpus to suppress background content. As a result, for each original image, we got a corresponding background-suppressed image across the whole corpus.

\subsection{Recall and Average Matched Similarity}\label{metrics:evaluation}
For every step where we compared the text from the ground-truth set with the OCR-generated text either from original images, background-suppressed images, or from different phases of the framework, we compared them using the following approach.

Given a manually-transcribed string $s_{gt}$ and an OCR-extracted string $s_{ocr}$, we compute their similarity as:

\begin{equation}
\label{eq:similarity}
\mathrm{Sim}(s_{gt}, s_{ocr}) = \frac{2 \cdot M(s_{gt}, s_{ocr})}{|s_{gt}| + |s_{ocr}|}
\end{equation}

where $|s_{gt}|$ and $|s_{ocr}|$ denote the lengths of the strings, and $M(s_{gt}, s_{ocr})$ represents the number of matching characters between the two strings.

A match between the two strings is determined using a similarity threshold $\tau$:

\begin{equation}
\label{eq:match}
\mathrm{Match}(s_{gt}, s_{ocr}) =
\begin{cases}
1, & \text{if } \mathrm{Sim}(s_{gt}, s_{ocr}) \ge \tau \\
0, & \text{otherwise}
\end{cases}
\end{equation}

In our evaluation, we set $\tau = 0.8$, allowing for minor OCR errors such as character substitutions, spacing inconsistencies, and punctuation differences. A stricter threshold (e.g., $\tau = 1.0$) would require exact string matches and would incorrectly penalize otherwise accurate OCR outputs that differ only in minor formatting or recognition errors.

\textit{AMS (Average Matched Similarity)} is computed as the average similarity over all successfully matched pairs:

\begin{equation}
\label{eq:ams}
\mathrm{AMS} = \frac{1}{\sum_{i=1}^{N} \mathrm{Match}(s_{gt}^{(i)}, s_{ocr}^{(i)})}
\sum_{i=1}^{N} \mathrm{Match}(s_{gt}^{(i)}, s_{ocr}^{(i)}) \cdot \mathrm{Sim}(s_{gt}^{(i)}, s_{ocr}^{(i)})
\end{equation}

\textit{Recall} is computed as the proportion of ground-truth lines that are successfully matched to an OCR-extracted line:

\begin{equation}
\label{eq:accuracy}
\mathrm{Recall} = \frac{1}{N} \sum_{i=1}^{N} \mathrm{Match}(s_{gt}^{(i)}, s_{ocr}^{(i)})
\end{equation}

where $N$ is the total number of ground-truth lines.

\subsection{LLM Selection}\label{subsec:llm:selection}
To select an LLM with reasonable performance, we evaluate the performance of several state-of-the-art LLMs on a manually curated ground truth dataset. Here, we discuss our dataset annotation process and our evaluation results.

\noindent
\textbf{Data Annotation.} To create the ground truth dataset, we randomly sampled 2,000 conversations across games and age categories, which were labeled by two annotators. Both annotators had prior familiarity with online safety risks on youth-oriented platforms, including exposure to common indicators of grooming, harassment, and inappropriate content through academic coursework, research experience in content moderation, and engagement with relevant online platforms. 
Disagreements were resolved through discussion. We restricted the annotation and analysis to English conversations (33 out of 2000 conversations were non-English). Finally, we labeled 1,680 conversations as \textit{safe} and 287 conversations as \textit{unsafe}, reflecting the relative rarity of unsafe interactions; it also creates a realistic \textit{imbalanced} evaluation setting in which models must reliably identify a minority class of high importance. 





\noindent
\textbf{LLM Selection and Prompting Methods.} \label{sec:expwithmodels}
Using the ground truth dataset described above, we evaluated four open-weight models: Mistral 7B, Llama4, Phi-4, and gpt-oss-120b. 
All models were deployed locally using Ollama to avoid uploading messages to third-party cloud servers. For evaluation, we used both \textit{zero-shot} and \textit{few-shot} prompting approaches.
For zero-shot, the models were prompted to label one conversation at a time using the prompt shown in~\cref{fig:llm_prompt}. When a model's performance (precision, recall, and F1-score) was unsatisfactory, we analyzed cases where they diverged from the ground truth and identified systematic failure patterns. For example, grooming interactions were missed when they were expressed as trust-building. 
Based on this analysis, we selected examples representing false negatives, false positives, and representative correct classifications. For each example, we extracted a short conversation segment and aimed to capture a range of interaction patterns. These examples were then incorporated into the base prompt to guide few-shot prompting.

\begin{table}[H]
\centering
\caption{Evaluation results of candidate LLMs with zero- and few-shot prompting strategy in percentage. Here, accuracy means how often the LLM correctly labels safe or unsafe messages, and precision means how often the LLM marks a conversation as unsafe that is actually unsafe, whereas Recall means how often the LLM correctly labels it as unsafe.}
\label{tab:llm_validation}
\resizebox{\columnwidth}{!}{%
\begin{tabular}{lccccr}
\toprule
\textbf{Model} & \textbf{Accuracy} & \textbf{Precision} & \textbf{Recall} & \textbf{F1} & \textbf{Prompt-Strategy} \\
\midrule
gpt-oss-120B   & 69.48 & 26.09 & 73.46 & 38.51 & zero-shot \\
gpt-oss-120B   & 74.04 & 31.93 & 71.43 & 44.13 & few-shot \\
Mistral 7B & 66.23 & 20.34 & 46.34 & 28.27 & zero-shot \\ 
Mistral 7B & 43.87 & 18.63 & 86.41 & 30.66 & few-shot \\
Llama4  & 41.22 & 17.26 & 81.53 & 28.48 & zero-shot \\
Llama4  & 64.88 & 22.59 & 59.58 & 32.76 & few-shot \\
Phi-4    & 79.40 & 32.12 & 40.07 & 35.66 & zero-shot \\
Phi-4    & 80.69 & 35.48 & 42.16 & 38.54 & few-shot \\
\bottomrule
\end{tabular}
}
\end{table}
\noindent
\textbf{Results.} We adopt a \textit{four-label} schema (as defined in~\cref{fig:llm_prompt}) during inference to capture varying degrees of safety and model uncertainty (e.g., borderline or vague cases), which are common in informal and context-dependent chat. This approach allows the model to better express nuanced judgments and reduces forced binary decisions during classification. As the models produced outputs across four labels, we map ``Absolutely SAFE'' to the \textit{safe} class and the remaining labels to \textit{unsafe} for binary evaluation. This mapping reflects a design choice, where any indication of potential risk is treated as unsafe, aligning with the goal of minimizing missed harmful content. Implementing this approach, we compute standard evaluation metrics, including accuracy, precision, recall, and F1-score. As~\cref{tab:llm_validation} shows, \textit{gpt-oss-120b} with few-shot prompting achieves the most balanced performance across Accuracy and recall, resulting in the most robust overall F1-score. 

\subsection{Hash Span Length Distribution}\label{hash:length}

\begin{figure}[H]
    \centering
    \makebox[\linewidth][c]{%
        \includegraphics[width=1\linewidth]{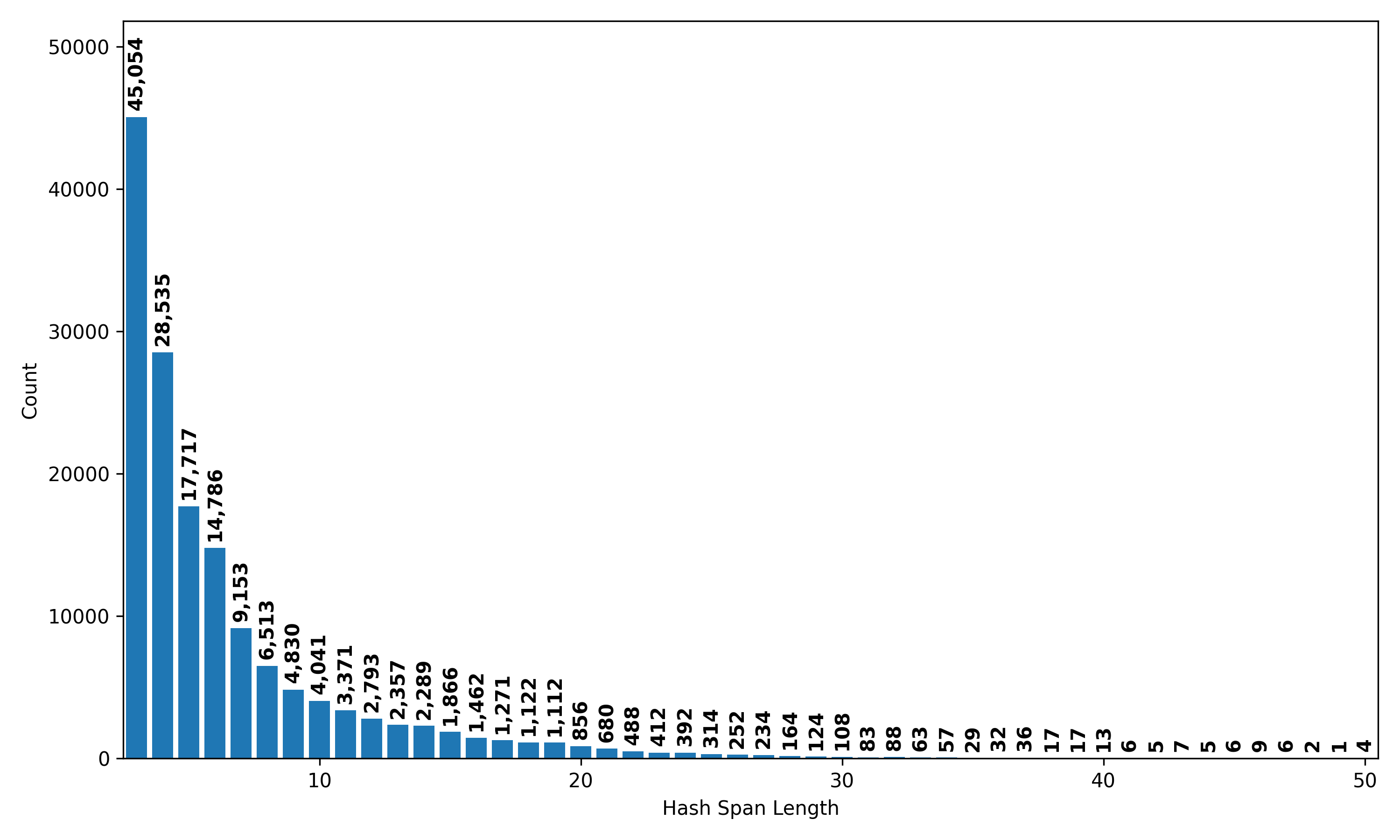}
    }
    \caption{Distribution of hash span lengths less than 50 characters long}
    \label{fig:hash-length-distribution}
\end{figure}

\subsection{LLM Prompt}\label{llm:prompt}

\begin{figure*}[h]
\centering
\footnotesize
\begin{tcolorbox}[
    colback=white,
    colframe=black,
    boxrule=0.6pt,
    arc=0mm,
    width=\textwidth,
    left=1mm,
    right=1mm,
    top=1mm,
    bottom=1mm,
    colbacktitle=black,
    coltitle=white,
    fonttitle=\bfseries,
    title=Standardized Prompt for LLM Classification.
]
\noindent\textbf{Role:} You are a content safety classifier trained to detect whether text conversations are unsafe for minors under the age of 18. Your goal is to classify the conversation into one of four categories:

\vspace{2mm}

\noindent\textbf{Absolutely UNSAFE:}

Clear and strong unsafe behavior, including:
\begin{itemize}\setlength{\itemsep}{1pt}\setlength{\parskip}{0pt}\setlength{\parsep}{0pt}
    \item[-] Sexual content, sexual innuendo, like talking about gay, lesbian, or any kind of sexuality
    \item[-] Explicit or mild flirting (Example: Want sum tonight, wanna go to bed, etc.)
    \item[-] Grooming behaviors (trust building, isolation, secrecy), including asking for personal information like age or birthdays
    \item[-] Attempts to move to another platform (YouTube, Instagram, Discord, Snapchat, etc.) includng saying usernames
    \item[-] Asking for age, gender, photos, or private contact information with suspicious intent
    \item[-] Explicit threats, harassment, hate speech, or discrimination or bullying by insulting or bad-mouthing someone
    \item[-] Encouraging self-harm, drugs, or alcohol use
    \item[-] Mild profanity without harmful context (e.g., "damn", "hell", "WTH", Ass/*ss/A\$\$/ASSI, f4 (f-word), btc (bitch))
\end{itemize}

\vspace{2mm}
\noindent\textbf{Possibly UNSAFE:}

Ambiguous but concerning signals, including:

\begin{itemize}\setlength{\itemsep}{1pt}\setlength{\parskip}{0pt}\setlength{\parsep}{0pt}
    \item[-] Mild flirting without clear sexual intent
    \item[-] Asking for age, gender, photos, or private contact information with suspicious intent
    \item[-] Context that may indicate grooming, but not explicit
\end{itemize}

\vspace{2mm}
\noindent\textbf{Possibly SAFE:}

Ambiguous but likely harmless:

\begin{itemize}\setlength{\itemsep}{1pt}\setlength{\parskip}{0pt}\setlength{\parsep}{0pt}
    \item[-] Compliments without romantic/sexual intent
    \item[-] Single personal question without follow-up
    \item[-] Short unclear phrases
    \item[-] Casual joking
\end{itemize}

\vspace{2mm}
\noindent\textbf{Absolutely SAFE:}

Clearly harmless conversation, including:

\begin{itemize}\setlength{\itemsep}{1pt}\setlength{\parskip}{0pt}\setlength{\parsep}{0pt}
    \item[-] Normal game discussion like "asking a charcater it's location in the gaming environment" or "asking a character to follow them in the game."
    \item[-] Friendly banter without inappropriate intent
    \item[-] Roleplay like "mom", "dad", "brother", "sister" that mimics family(Example: "Can you be my dad!")
    \item[-] Ambiguous Statements
    \item[-] Noisy OCR text that is unreadable and garbled
    \item[-] Non-English text
\end{itemize}

\vspace{2mm}
\noindent\textbf{IMPORTANT RULES:}
\begin{itemize}\setlength{\itemsep}{1pt}\setlength{\parskip}{0pt}\setlength{\parsep}{0pt}
    \item[-] Random OCR noise is SAFE
    \item[-] Non-English text is SAFE
    \item[-] Do NOT mark unsafe unless there is a clear risk
    \item[-] Focus on intent, not isolated words
\end{itemize}

\vspace{2mm}
\noindent\textbf{OUTPUT FORMAT (STRICT):}

Return EXACTLY two lines:

Decision: <Absolutely SAFE | Possibly SAFE | Possibly UNSAFE | Absolutely UNSAFE> \\
Reason: <brief explanation referencing key behavior or pattern>

\vspace{2mm}
Now classify this conversation:

Conversation:  
\texttt{<conversation\_text>}

\end{tcolorbox}
\caption{Standardized prompt used for LLM-based classification of conversations.}
\label{fig:llm_prompt}
\end{figure*}

\clearpage

%% file: bibliography.bib
@misc{clement2025roblox,
  title = {Roblox Global Daily Active Users},
  author = {Jessica Clement},
  year = {2025},
  howpublished={\url{https://www.statista.com/statistics/1192573/daily-active-users-global-roblox/}},
}

@misc{chatbot_app,
  author       = {{Chatbot App}},
  title        = {{AI Chatbot}},
  howpublished = {\url{https://chat.chatbot.app/?model=gemini-3-pro}},
  year         = {2026},
  note         = {Accessed: 2026-04-30}
}

@misc{chatgpt,
  author       = {{OpenAI}},
  title        = {{ChatGPT}},
  howpublished = {\url{https://chatgpt.com/}},
  year         = {2026},
  note         = {Accessed: 2026-04-30}
}

@misc{grammarly,
  author = {{Grammarly, Inc.}},
  title = {Grammarly},
  year = {2026},
  url = {https://app.grammarly.com/},
  note = {Accessed: 30 April 2026}
}

@article{DBLP:journals/corr/abs-2002-12543,
  author       = {Tsong Yueh Chen and
                  Shing{-}Chi Cheung and
                  Siu{-}Ming Yiu},
  title        = {Metamorphic Testing: {A} New Approach for Generating Next Test Cases},
  journal      = {CoRR},
  volume       = {abs/2002.12543},
  year         = {2020},
  url          = {https://arxiv.org/abs/2002.12543},
  eprinttype   = {arXiv},
  eprint       = {2002.12543}
}

@inproceedings{DBLP:conf/icwsm/FieslerBK20,
  author       = {Casey Fiesler and
                  Nathan Beard and
                  Brian C. Keegan},
  editor       = {Munmun De Choudhury and
                  Rumi Chunara and
                  Aron Culotta and
                  Brooke Foucault Welles},
  title        = {No Robots, Spiders, or Scrapers: Legal and Ethical Regulation of Data
                  Collection Methods in Social Media Terms of Service},
  booktitle    = {Proceedings of the Fourteenth International {AAAI} Conference on Web
                  and Social Media, {ICWSM} 2020, Held Virtually, Original Venue: Atlanta,
                  Georgia, USA, June 8-11, 2020},
  pages        = {187--196},
  publisher    = {{AAAI} Press},
  year         = {2020}
}

@inproceedings{DBLP:conf/sp/ThomasKTMVTBFEB25,
  author       = {Kurt Thomas and
                  Patrick Gage Kelley and
                  David Tao and
                  Sarah Meiklejohn and
                  Owen Vallis and
                  Shunwen Tan and
                  Blaz Bratanic and
                  Felipe Tiengo Ferreira and
                  Vijay Kumar Eranti and
                  Elie Bursztein},
  editor       = {Marina Blanton and
                  William Enck and
                  Cristina Nita{-}Rotaru},
  title        = {Supporting Human Raters with the Detection of Harmful Content Using
                  Large Language Models},
  booktitle    = {{IEEE} Symposium on Security and Privacy, {SP} 2025, San Francisco,
                  CA, USA, May 12-15, 2025},
  pages        = {2772--2789},
  publisher    = {{IEEE}},
  year         = {2025}
}

@misc{Koneru_2025, 
  title={How Roblox uses AI to Moderate Content on a Massive Scale}, 
  url={https://about.roblox.com/newsroom/2025/07/roblox-ai-moderation-massive-scale}, 
  journal={Roblox}, 
  author={Koneru, Naren}, 
  year={2025}, 
  month={Jul}
}

@misc{robloxsafety,
  author       = {Roblox Corporation},
  title        = {Roblox Safety Tools},
  year         = {2026},
  howpublished = {\url{https://about.roblox.com/safety-tools}}
}

@article{eskicioglu1995image,
  title={Image quality measures and their performance},
  author={Eskicioglu, Ahmet M and Fisher, Paul S},
  journal={IEEE Transactions on communications},
  volume={43},
  number={12},
  pages={2959--2965},
  year={1995},
  publisher={IEEE}
}

@misc{LACountyRoblox2026,
  title        = {People v. Roblox},
  author       = {{People of the State of California, ex rel. Los Angeles County Counsel}},
  year         = {2026},
  note         = {Filed: February 19, 2026},
  howpublished = {\url{https://file.lacounty.gov/SDSInter/lac/1202405_Complaint-Peoplev.Roblox.pdf}},
}

@article{doring2014consensual,
  title={Consensual sexting among adolescents: Risk prevention through abstinence education or safer sexting?},
  author={D{\"o}ring, Nicola},
  journal={Cyberpsychology: Journal of Psychosocial Research on Cyberspace},
  volume={8},
  number={1},
  year={2014}
}

@misc{seitz,
  author = {US Distric Court Eastern District of Kentucky},
  title = {Seitz v. Roblox},
  howpublished = {\url{https://www.anapolweiss.com/2025-10-20-complaint.pdf}},
  year = {2025},
  note = {Filed: October 20, 2025}
}

@misc{Carville_DAnastasio_2024, 
  title={Roblox’s Pedophile Problam}, 
  url={https://www.bloomberg.com/features/2024-roblox-pedophile-problem/}, 
  journal={Bloomberg}, 
  author={Carville, Olivia and D’Anastasio, Cecilia}, 
  year={2024}
}

@misc{Ellery_2025, 
  title={Roblox Safety Failings Leave Children at Risk, Claim Experts}, 
  url={https://www.thetimes.com/uk/law/article/roblox-kk-swastikas-childrens-safety-fx5n6tcl7}, 
  journal={The Times}, 
  author={Ellery, Ben}, 
  year={2025}
}

@misc{Chawki_2025, 
  title={AI Moderation and Legal Frameworks in Child-Centric Social Media: A Case Study of Roblox}, 
  url={https://doi.org/10.3390/laws14030029}, 
  journal={MDPI}, 
  author={Chawki, Mohamed}, 
  year={2025}
}

@article{mudeng2022prospects,
  title={Prospects of structural similarity index for medical image analysis},
  author={Mudeng, Vicky and Kim, Minseok and Choe, Se-woon},
  journal={Applied Sciences},
  volume={12},
  number={8},
  pages={3754},
  year={2022},
  publisher={MDPI}
}

@article{chang2015ssim,
  title={SSIM-based quality-on-demand energy-saving schemes for OLED displays},
  author={Chang, Teng-Chang and Xu, Sendren Sheng-Dong and Su, Shun-Feng},
  journal={IEEE Transactions on Systems, Man, and Cybernetics: Systems},
  volume={46},
  number={5},
  pages={623--635},
  year={2015},
  publisher={IEEE}
}

@article{wang2004image,
  title={Image quality assessment: from error visibility to structural similarity},
  author={Wang, Zhou and Bovik, Alan C and Sheikh, Hamid R and Simoncelli, Eero P},
  journal={IEEE transactions on image processing},
  volume={13},
  number={4},
  pages={600--612},
  year={2004},
  publisher={IEEE}
}

@book{villan2019mastering,
  title={Mastering OpenCV 4 with Python: a practical guide covering topics from image processing, augmented reality to deep learning with OpenCV 4 and Python 3.7},
  author={Vill{\'a}n, Alberto Fern{\'a}ndez},
  year={2019},
  publisher={Packt Publishing Ltd},
  address={Birmingham, UK}
}

@misc{sweigart2020pyautogui,
  title = {PyAutoGUI: Cross-platform GUI automation for Python},
  author = {Al Sweigart},
  year = {2025},
  url = {https://github.com/asweigart/pyautogui}
}

@INPROCEEDINGS{10190527,
  author={Singhal, Mohit and Ling, Chen and Paudel, Pujan and Thota, Poojitha and Kumarswamy, Nihal and Stringhini, Gianluca and Nilizadeh, Shirin},
  booktitle={2023 IEEE 8th European Symposium on Security and Privacy (EuroS\&P)}, 
  title={SoK: Content Moderation in Social Media, from Guidelines to Enforcement, and Research to Practice}, 
  year={2023},
  pages={868-895},
  doi={10.1109/EuroSP57164.2023.00056},
  publisher={IEEE},
  address={Delft, Netherlands}
}

@article{flynn2024child,
  title={Child protection and welfare risks and opportunities related to disability and internet use: Broadening current conceptualisations through critical literature review},
  author={Flynn, Susan and Maher, Rose Doolan and Byrne, Julie},
  journal={Children and Youth Services Review},
  volume={157},
  pages={107410},
  year={2024},
  publisher={Elsevier}
}

@book{livingstone2012children,
  title={Children, risk and safety on the internet: Research and policy challenges in comparative perspective},
  author={Livingstone, Sonia and Haddon, Leslie},
  year={2012},
  publisher={Policy Press},
  address={Chicago, IL, USA}
}

@inproceedings{goyal2025momoe,
  title={MoMoE: Mixture of Moderation Experts Framework for AI-Assisted Online Governance},
  author={Goyal, Agam and Zhan, Xianyang and Chen, Yilun and Saha, Koustuv and Chandrasekharan, Eshwar},
  booktitle={Proceedings of the 2025 Congerence on Empirical Methods in Natural Language Processing},
  pages = {12645--12660},
  publisher = {Association for Computational Linguistics},
  url = {https://aclanthology.org/2025.emnlp-main.638/},
  doi = {10.18653/v1/2025.emnlp-main.638},
  address = {Suzhou, China},
  year={2025}
}

@misc{eltaher2025protecting,
  title={Protecting Young Users on Social Media: Evaluating the Effectiveness of Content Moderation and Legal Safeguards on Video Sharing Platforms},
  author={Eltaher, Fatmaelzahraa and Gajula, Rahul Krishna and Miralles-Pechu{\'a}n, Luis and Crotty, Patrick and Mart{\'\i}nez-Otero, Juan and Thorpe, Christina and McKeever, Susan},
  journal={arXiv preprint arXiv:2505.11160},
  year={2025}
}

@article{yousaf2022deep,
  title={A deep learning-based approach for inappropriate content detection and classification of youtube videos},
  author={Yousaf, Kanwal and Nawaz, Tabassam},
  journal={IEEE Access},
  volume={10},
  pages={16283--16298},
  year={2022},
  publisher={IEEE}
}

@article{an2025toward,
  title={Toward Integrated Solutions: A Systematic Interdisciplinary Review of Cybergrooming Research},
  author={An, Heajun and Silva, Marcos and Zhang, Qi and Singh, Arav and Liu, Minqian and Zhang, Xinyi and Qadir, Sarvech and Lee, Sang Won and Huang, Lifu and Wisnieswski, Pamela and others},
  journal={arXiv preprint arXiv:2503.05727},
  year={2025}
}

@article{finkelhor2022prevalence,
  title={Prevalence of online sexual offenses against children in the US},
  author={Finkelhor, David and Turner, Heather and Colburn, Deirdre},
  journal={JAMA network open},
  volume={5},
  number={10},
  pages={e2234471--e2234471},
  year={2022},
  publisher={American Medical Association}
}

@misc{bihani2025fuzzy,
  title={A Fuzzy Evaluation of Sentence Encoders on Grooming Risk Classification}, 
  author={Geetanjali Bihani and Julia Rayz},
  year={2025},
  eprint={2502.12576},
  archivePrefix={arXiv},
  primaryClass={cs.CL},
  url={https://arxiv.org/abs/2502.12576}, 
}

@article{whittle2013review,
  title={A review of online grooming: Characteristics and concerns},
  author={Whittle, Helen and Hamilton-Giachritsis, Catherine and Beech, Anthony and Collings, Guy},
  journal={Aggression and violent behavior},
  volume={18},
  number={1},
  pages={62--70},
  year={2013},
  publisher={Elsevier}
}

@article{engelmann2025developing,
  title={Developing quality standards for community-based online child sexual exploitation and abuse interventions},
  author={Engelmann, Larissa and Weirich, Christine A and May-Chahal, Corinne},
  journal={Child Abuse \& Neglect},
  volume={164},
  pages={107444},
  year={2025},
  publisher={Elsevier}
}

@article{aliyeva2025deep,
  title={Deep learning approach to detect cyberbullying on twitter},
  author={Aliyeva, {\c{C}}inare O{\u{g}}uz and Ya{\u{g}}ano{\u{g}}lu, Mete},
  journal={Multimedia Tools and Applications},
  volume={84},
  number={19},
  pages={20497--20520},
  year={2025},
  publisher={Springer}
}

@article{mccrae2017social,
  title={Social media and depressive symptoms in childhood and adolescence: A systematic review},
  author={McCrae, Niall and Gettings, Sheryl and Purssell, Edward},
  journal={Adolescent Research Review},
  volume={2},
  number={4},
  pages={315--330},
  year={2017},
  publisher={Springer}
}

@article{hinduja2013social,
  title={Social influences on cyberbullying behaviors among middle and high school students},
  author={Hinduja, Sameer and Patchin, Justin W},
  journal={Journal of youth and adolescence},
  volume={42},
  number={5},
  pages={711--722},
  year={2013},
  publisher={Springer}
}

@article{radesky2024algorithmic,
  title={Algorithmic content recommendations on a video-sharing platform used by children},
  author={Radesky, Jenny and Bridgewater, Enrica and Black, Shira and O’Neil, August and Sun, Yilin and Schaller, Alexandria and Weeks, Heidi M and Campbell, Scott W},
  journal={JAMA Network Open},
  volume={7},
  number={5},
  pages={e2413855--e2413855},
  year={2024},
  publisher={American Medical Association}
}

@article{pasquale2020digital,
  title={Digital age of consent and age verification: can they protect children?},
  author={Pasquale, Liliana and Zippo, Paola and Curley, Cliona and O’Neill, Brian and Mongiello, Marina},
  journal={IEEE software},
  volume={39},
  number={3},
  pages={50--57},
  year={2020},
  publisher={IEEE}
}

@book{szeliski2022computer,
  title={Computer vision: algorithms and applications},
  author={Szeliski, Richard},
  year={2022},
  publisher={Springer Nature}
}

@misc{coppareg,
  author       = {Federal Trade Commission},
  title        = {Children's Online Privacy Protection Rule: Final Rule Amendments},
  howpublished = {16 C.F.R. Part 312},
  year         = {2013},
  note         = {78 FR 3972},
  url          = {https://www.ftc.gov/legal-library/browse/rules/childrens-online-privacy-protection-rule-coppa}
}

@article{proferes2021studying,
  title={Studying reddit: A systematic overview of disciplines, approaches, methods, and ethics},
  author={Proferes, Nicholas and Jones, Naiyan and Gilbert, Sarah and Fiesler, Casey and Zimmer, Michael},
  journal={Social media+ society},
  volume={7},
  number={2},
  pages={20563051211019004},
  year={2021},
  publisher={SAGE Publications Sage UK: London, England}
}

@article{wickstrom2000hawthorne,
  title={The" Hawthorne effect"—what did the original Hawthorne studies actually show?},
  author={Wickstr{\"o}m, Gustav and Bendix, Tom},
  journal={Scandinavian journal of work, environment \& health},
  pages={363--367},
  year={2000},
  publisher={JSTOR}
}

@article{hudson2004go,
  title={“Go away”: Participant objections to being studied and the ethics of chatroom research},
  author={Hudson, James M and Bruckman, Amy},
  journal={The information society},
  volume={20},
  number={2},
  pages={127--139},
  year={2004},
  publisher={Taylor \& Francis}
}

@misc{robloxcomm,
  author       = {Roblox Corporation},
  title        = {Roblox Community Standards
},
  howpublished = {\url{https://en.help.roblox.com/hc/en-us/articles/203313410-Roblox-Community-Standards}},
  year = {2026}
}

@misc{robloxRobloxTerms,
	author = {Roblox Corporation},
	title = {{R}oblox {T}erms of {U}se --- en.help.roblox.com},
	howpublished = {\url{https://en.help.roblox.com/hc/en-us/articles/115004647846-{R}oblox-{T}erms-of-{U}se}},
	year = {2026}
}

@misc{robloxRobloxCommunity,
	author = {Roblox Corporation},
	title = {{R}oblox {C}ommunity {S}tandards | {R}oblox --- about.roblox.com},
	howpublished = {\url{https://about.roblox.com/community-standards}},
	year = {2026},
}

@article{stano2023linguistic,
  title={Linguistic guerrilla warfare 2.0: On the “forms” of online resistance},
  author={Stano, Simona and others},
  journal={Rivista Italiana di Filosofia del Linguaggio},
  volume={2022},
  pages={177--186},
  year={2023}
}

@inproceedings{kou2025system,
  title={“The System is Made to Inherently Push Child Gambling in my Opinion”: Child Safety, Monetization, and Moderation on Roblox},
  author={Kou, Yubo and Hernandez, Rie Helene and Gui, Xinning},
  booktitle={Proceedings of the 2025 CHI Conference on Human Factors in Computing Systems},
  pages={1--18},
  year={2025}
}

@article{craven2006sexual,
  title={Sexual grooming of children: Review of literature and theoretical considerations},
  author={Craven, Samantha and Brown, Sarah and Gilchrist, Elizabeth},
  journal={Journal of sexual aggression},
  volume={12},
  number={3},
  pages={287--299},
  year={2006},
  publisher={Taylor \& Francis}
}

@misc{opencvOpenCVImage,
	author = {Bradski, G},
	title = {{O}pen{C}{V}: {I}mage {T}hresholding --- docs.opencv.org},
	howpublished = {\url{https://docs.opencv.org/4.x/d7/da8/tutorial_table_of_content_imgproc.html}},
	year = {2025},
	note = {[Accessed 19-03-2026]},
}

@article{DBLP:conf/icmla/AhmedHS23,
  title={The Potential of Vision-Language Models for Content Moderation of Children's Videos},
  author={Syed Hammad Ahmed and Shengnan Hu and Gita Reese Sukthankar},
  journal={2023 International Conference on Machine Learning and Applications (ICMLA)},
  year={2023},
  pages={1237-1241},
  url={https://api.semanticscholar.org/CorpusID:266053028}
}

@inproceedings{zhao2019make,
  title={I make up a silly name' Understanding Children's Perception of Privacy Risks Online},
  author={Zhao, Jun and Wang, Ge and Dally, Carys and Slovak, Petr and Edbrooke-Childs, Julian and Van Kleek, Max and Shadbolt, Nigel},
  booktitle={Proceedings of the 2019 CHI conference on human factors in computing systems},
  publisher = {Association for Computing Machinery},
  address = {New York, NY, USA},
  doi = {10.1145/3290605.3300336},
  pages = {1–13},
  numpages = {13},
  year={2019},
  articleno={106}
}

@article{mrug2016emotional,
  title={Emotional desensitization to violence contributes to adolescents’ violent behavior},
  author={Mrug, Sylvie and Madan, Anjana and Windle, Michael},
  journal={Journal of abnormal child psychology},
  volume={44},
  number={1},
  pages={75--86},
  year={2016},
  publisher={Springer}
}

@article{veseli2025positional,
  title={Positional Biases Shift as Inputs Approach Context Window Limits},
  author={Veseli, Blerta and Chibane, Julian and Toneva, Mariya and Koller, Alexander},
  journal={arXiv preprint arXiv:2508.07479},
  year={2025}
}

@article{razi2023sliding,
  title={Sliding into my DMs: Detecting uncomfortable or unsafe sexual risk experiences within Instagram direct messages grounded in the perspective of youth},
  author={Razi, Afsaneh and Alsoubai, Ashwaq and Kim, Seunghyun and Ali, Shiza and Stringhini, Gianluca and De Choudhury, Munmun and Wisniewski, Pamela J},
  journal={Proceedings of the ACM on human-computer interaction},
  volume={7},
  number={CSCW1},
  pages={1--29},
  year={2023},
  publisher={ACM New York, NY, USA}
}

@article{agarwal2025gpt,
  title={gpt-oss-120b \& gpt-oss-20b model card},
  author={Agarwal, Sandhini and Ahmad, Lama and Ai, Jason and Altman, Sam and Applebaum, Andy and Arbus, Edwin and Arora, Rahul K and Bai, Yu and Baker, Bowen and Bao, Haiming and others},
  journal={arXiv preprint arXiv:2508.10925},
  year={2025}
}

@article{gonccalves2026potential,
  title={Potential Exposure of Kids to Age-Inappropriate Content on Twitch: A Comparative Cross-Country Study},
  author={Gon{\c{c}}alves, K{\^e}nia C and Soriano, Fl{\'a}vio and Marques-Neto, Humberto T and Almeida, Jussara M},
  journal={Social Network Analysis and Mining},
  volume={16},
  number={1},
  pages={1},
  year={2026},
  publisher={Springer}
}

@misc{robloxparental,
  author       = {Roblox Corporation},
  title        = {Parental Controls Overview},
  howpublished = {\url{https://en.help.roblox.com/hc/en-us/articles/30428310121620-Parental-Controls-Overview}},
  year={2025}
}

@misc{moderation,
  author       = {Roblox Corporation},
  title        = {Content Moderation on Roblox},
  howpublished = {\url{https://en.help.roblox.com/hc/en-us/articles/21416271342868-Content-Moderation-on-Roblox}},
  year = {2026}
}

@misc{instagramparental,
  author       = {Instagram},
  title        = {Parental Supervision},
  howpublished = {\url{https://help.instagram.com/309877544512275}},
  year={2025}
}

@article{hong2024impact,
  title={The impact of social media in child sexual abuse},
  author={Hong, Monica},
  journal={Journal of Paediatrics and Child Health},
  volume={60},
  number={10},
  pages={476--478},
  year={2024},
  publisher={Wiley Online Library}
}

@article{yang2025realfactbench,
  title={RealFactBench: A Benchmark for Evaluating Large Language Models in Real-World Fact-Checking},
  author={Shuo Yang and Yuqin Dai and Guoqing Wang and Xinran Zheng and Jinfeng Xu and Jinze Li and Zhenzhe Ying and Weiqiang Wang and Edith C. H. Ngai},
  journal={Proceedings of the 33rd ACM International Conference on Multimedia},
  year={2025},
  url={https://api.semanticscholar.org/CorpusID:279403389},
  pages = {13435–13441},
}

@misc{qureshi2025explainable,
  author = {Qureshi, Muhammad Deedahwar Mazhar and Qureshi, M Atif and Rashwan, Wael},
  title = {Explainable AI for Hate Speech Moderation: A Stakeholder-Centered and Socially Grounded Review},
  year = {2025},
  month = {aug},
  publisher = {TechRxiv},
  doi = {10.36227/techrxiv.175440435.54783623/v1},
  url = {https://www.techrxiv.org/users/948947/articles/1319625-explainable-ai-for-hate-speech-moderation-a-stakeholder-centered-and-socially-grounded-review}
}

@article{makhijani2021quest,
  title={Quest: Queue simulation for content moderation at scale},
  author={Makhijani, Rahul and Shah, Parikshit and Avadhanula, Vashist and Gocmen, Caner and Stier-Moses, Nicol{\'a}s E and Mestre, Juli{\'a}n},
  journal={arXiv preprint arXiv:2103.16816},
  year={2021},
  volume={2103.16816},
  numpages={9}
}

@article{gillespie2020content,
  title={Content moderation, AI, and the question of scale},
  author={Gillespie, Tarleton},
  journal={Big Data \& Society},
  volume={7},
  number={2},
  pages={2053951720943234},
  year={2020},
  publisher={SAGE Publications Sage UK: London, England}
}

@inproceedings{10.1145/3406865.3418312,
author = {Jiang, Jialun 'Aaron' and Middler, Skyler and Brubaker, Jed R. and Fiesler, Casey},
title = {Characterizing Community Guidelines on Social Media Platforms},
year = {2020},
isbn = {9781450380591},
publisher = {Association for Computing Machinery},
address = {New York, NY, USA},
url = {https://doi.org/10.1145/3406865.3418312},
doi = {10.1145/3406865.3418312},
booktitle = {Companion Publication of the 2020 Conference on Computer Supported Cooperative Work and Social Computing},
pages = {287–291},
numpages = {5},
location = {Virtual Event, USA},
series = {CSCW '20 Companion}
}

@article{seering2019moderator,
  title={Moderator engagement and community development in the age of algorithms},
  author={Seering, Joseph and Wang, Tony and Yoon, Jina and Kaufman, Geoff},
  journal={New media \& society},
  volume={21},
  number={7},
  pages={1417--1443},
  year={2019},
  publisher={SAGE Publications Sage UK: London, England}
}

@inproceedings{cai2024content,
  title={Content moderation justice and fairness on social media: Comparisons across different contexts and platforms},
  author={Cai, Jie and Patel, Aashka and Naderi, Azadeh and Wohn, Donghee Yvette},
  booktitle={Extended Abstracts of the CHI Conference on Human Factors in Computing Systems},
  numpages={9},
  articleno = {84},
  isbn = {9798400703317},
  publisher = {Association for Computing Machinery},
  address = {New York, NY, USA},
  url = {https://doi.org/10.1145/3613905.3650882},
  doi = {10.1145/3613905.3650882},
  year={2024}
}

@inproceedings{annamoradnejad2022requirements,
  title={Requirements for automating moderation in community question-answering websites},
  author={Annamoradnejad, Issa},
  booktitle={Proceedings of the 15th Innovations in Software Engineering Conference},
  pages={1--4},
  year={2022},
  publisher = {Association for Computing Machinery},
  address = {New York, NY, USA},
}

@inproceedings{gatta2023interconnected,
  title={The interconnected nature of online harm and moderation: Investigating the cross-platform spread of harmful content between youtube and twitter},
  author={Gatta, Valerio La and Luceri, Luca and Fabbri, Francesco and Ferrara, Emilio},
  booktitle={Proceedings of the 34th ACM conference on hypertext and social media},
  pages={1--10},
  year={2023}
}

@article{goldstein2023understanding,
  title={Understanding the (in) effectiveness of content moderation: A case study of facebook in the context of the us capitol riot},
  author={Goldstein, Ian and Edelson, Laura and Nguyen, Minh-Kha and Goga, Oana and McCoy, Damon and Lauinger, Tobias},
  journal={arXiv preprint arXiv:2301.02737},
  year={2023}
}

@article{yang2021tar,
  title={TAR on social media: A framework for online content moderation},
  author={Yang, Eugene and Lewis, David D and Frieder, Ophir},
  journal={arXiv preprint arXiv:2108.12752},
  year={2021},
  pages={147--155},
  volume={2950}
}

@misc{de2021peer,
  title={Peer Governance in Online Communities},
  author={De Filippi, Primavera and Schneider, Nathan},
  journal={Frontiers in Human Dynamics},
  volume={3},
  pages={771586},
  year={2021},
  publisher={Frontiers Media SA}
}

@article{matias2019civic,
  title={The Civic Labor of Volunteer Moderators Online},
  author={Matias, J Nathan},
  journal={Social Media+ Society},
  volume={5},
  number={2},
  pages={2056305119836778},
  year={2019},
  publisher={SAGE Publications Sage UK: London, England}
}

@inproceedings{10.1145/3290605.3300390,
author = {Wohn, Donghee Yvette},
title = {Volunteer Moderators in Twitch Micro Communities: How They Get Involved, the Roles They Play, and the Emotional Labor They Experience},
year = {2019},
isbn = {9781450359702},
publisher = {Association for Computing Machinery},
address = {New York, NY, USA},
url = {https://doi.org/10.1145/3290605.3300390},
doi = {10.1145/3290605.3300390},
booktitle = {Proceedings of the 2019 CHI Conference on Human Factors in Computing Systems},
pages = {1–13},
numpages = {13},
articleno={160}
}

@article{kayany1998contexts,
  title={Contexts of Uninhibited Online Behavior: Flaming in Social Newsgroups on Usenet},
  author={Kayany, Joseph M},
  journal={Journal of the American Society for Information Science},
  volume={49},
  number={12},
  pages={1135--1141},
  year={1998},
  publisher={Wiley Online Library}
}

@INPROCEEDINGS{1241285,
  author={Arnt, A. and Zilberstein, S.},
  booktitle={Proceedings IEEE/WIC International Conference on Web Intelligence (WI 2003)}, 
  title={Learning to Perform Moderation in Online Forums}, 
  year={2003},
  publisher={IEEE},
  pages={637-641},
  doi={10.1109/WI.2003.1241285},
  address={Halifax, NS, Canada}
}

@article{Sravanti2025,
author = {Lakshmi Sravanti and Arul Jayendra Pradeep Velusamy and Kiragasur Madegowda Rajendra and John Vijay Sagar Kommu},
title ={Childhood Digital Exposure to Sexual Content: Through the Lens of Developmental Psychopathology},

journal = {Journal of Psychosexual Health},
volume = {7},
number = {2},
pages = {115-119},
year = {2025},
doi = {10.1177/26318318251322555},
URL = {https://doi.org/10.1177/26318318251322555}
}

@article{Ramadhani_Khodari_Ulfiah_Rosydawati_2025, 
  title={Silence Out of Fear: A Case Study of the Spiral of Silence in Social Media Doxxing Victims}, 
  volume={4}, 
  DOI={https://doi.org/10.58344/jmi.v4i6.2506}, 
  number={6}, 
  journal={Jurnal Multidisiplin Indonesia}, 
  author={Ramadhani, Sansa Aldira and Khodari, Rifqi Taufiqurrohman and Ulfiah, Naily and Rosydawati, Aprilia}, 
  year={2025}, 
  month={Dec}, 
  pages={417–428}
}

@misc{Tan_2025,
  title={He Made a Friend on Roblox. Their Relationship Turned Sinister.}, 
  url={https://www.nytimes.com/2025/09/12/technology/roblox-lawsuit-child-safety.html}, 
  journal={New York Times}, 
  author={Tan, Eli}, 
  year={2025}, 
  month={Sep}
}

@article{Dendi_2025, 
  title={Real-Time Content Moderation in Gaming Platforms: Technical Frameworks for Child Protection}, 
  volume={7}, 
  url={https://al-kindipublishers.org/index.php/jcsts/article/view/10735}, 
  DOI={10.32996/jcsts.2025.7.9.1}, 
  number={9}, 
  journal={Journal of Computer Science and Technology Studies}, 
  author={Naveen Reddy Dendi}, 
  year={2025}, 
  month={Aug.}, 
  pages={01–08} 
}

@article{Garcia_Carvalho_2025, 
  title={A Literature Review of Textual Cyber Abuse Detection Using Cutting‐Edge Natural Language Processing Techniques: Language Models and Large Language Models}, 
  volume={15}, 
  DOI={10.1002/widm.70029}, 
  number={3}, 
  journal={WIREs Data Mining and Knowledge Discovery}, 
  author={Diaz‐Garcia, J. Angel and Carvalho, Joao Paulo}, 
  year={2025}, 
  month={Jun},
  numpages={38}
}

@article{computation13080196,
  AUTHOR = {Barakat, Basel and Jaf, Sardar},
  TITLE = {Beyond Traditional Classifiers: Evaluating Large Language Models for Robust Hate Speech Detection},
  JOURNAL = {Computation},
  VOLUME = {13},
  YEAR = {2025},
  NUMBER = {8},
  articleno = {196},
  URL = {https://www.mdpi.com/2079-3197/13/8/196},
  ISSN = {2079-3197},
  numpages={19},
  DOI = {10.3390/computation13080196}
}

@misc{Statista_2025, 
  title={Gaming Reach Worldwide by Age and Gender 2025}, 
  url={https://www.statista.com/statistics/326420/console-gamers-gender/}, 
  journal={Statista}, 
  author={Statista Research Department}, 
  year={2025}, 
  month={Nov}
}

@article{OSG_Parcelli,
  author = {Carly Porcelli and Flora Anderson and Juliane A. Kloess},
  title ={Unraveling the Complexities of Offender Strategies as Part of Online Sexual Grooming and Technology-Assisted Child Sexual Abuse: A Systematic Review},
  journal = {Trauma, Violence, \& Abuse},
  volume = {0},
  number = {0},
  pages = {15248380251411261},
  year = {0},
  doi = {10.1177/15248380251411261},
  note ={PMID: 41618489},
  URL = {https://doi.org/10.1177/15248380251411261}
}

@misc{Wrocherinsky_2023, 
  title={Public Panics and Youth Online Safety – a Deep Dive}, 
  url={https://project-disco.org/featured/public-panics-and-youth-online-safety-a-deep-dive/}, 
  journal={Disruptive Competition Project}, 
  author={Maranon, Alvaro and Wrocherinsky, Dalia}, 
  year={2023}, 
  month={Jul}
}

@inbook{Hasan_Athrey_Khalid_Xie_Younessian_Braskich_2024, 
  chapter={Applications of computer vision in entertainment and Media Industry}, 
  DOI={10.1201/9781003328957-10}, 
  title={Computer Vision: Challenges, Trends, and Opportunies}, 
  author={Hasan, Mahmudul and Athrey, Kishan Shamsundar and Khalid, Arfeen and Xie, Danfeng and Younessian, Ehsan and Braskich, Tony}, 
  year={2024}, 
  month={May}, 
  pages={205–238},
  publisher={CRC Press},
  address={Boca Raton, FL}
}

@ARTICLE{10680313,
  author={Gallotta, Roberto and Todd, Graham and Zammit, Marvin and Earle, Sam and Liapis, Antonios and Togelius, Julian and Yannakakis, Georgios N.},
  journal={IEEE Transactions on Games}, 
  title={Large Language Models and Games: A Survey and Roadmap}, 
  year={2024},
  pages={1-18},
  doi={10.1109/TG.2024.3461510}
}

@article{Briskilal_2024, 
  title={Detection of offensive text in memes using Deep Learning Techniques}, 
  volume={3075}, 
  DOI={10.1063/5.0217063}, 
  journal={AIP Conference Proceedings}, 
  author={Briskilal, J. and Karthik, M. Jaya and Praneeth, Sai},
  pages={124484-124498},
  year={2024}, 
  month={Jul}
}

@Article{cmes.2025.061653,
AUTHOR = {Sangmin Kim and Byeongcheon Lee and Muazzam Maqsood and Jihoon Moon and Seungmin Rho},
TITLE = {Deep Learning-Based Natural Language Processing Model and Optical Character Recognition for Detection of Online Grooming on Social Networking Services},
JOURNAL = {Computer Modeling in Engineering \& Sciences},
VOLUME = {143},
YEAR = {2025},
NUMBER = {2},
PAGES = {2079--2108},
URL = {http://www.techscience.com/CMES/v143n2/61423},
ISSN = {1526-1506},
DOI = {10.32604/cmes.2025.061653}
}

@INPROCEEDINGS{11168676,
  author={Jose, Mekha and Anthony, Jocelyn and Joseph, Jose V and Thomas, Joshwa and Thomas, Sharon Baby},
  booktitle={2025 Advanced Computing and Communication Technologies for High Performance Applications (ACCTHPA)}, 
  title={Automated Detection of Offensive Text in Social Media Images}, 
  year={2025},
  publisher={IEEE},
  address={Cochin, Kerala, India},
  pages={1-6},
  doi={10.1109/ACCTHPA65749.2025.11168676}}

@inproceedings{Thomas_2025, 
  author       = {Kurt Thomas and
                  Patrick Gage Kelley and
                  David Tao and
                  Sarah Meiklejohn and
                  Owen Vallis and
                  Shunwen Tan and
                  Blaz Bratanic and
                  Felipe Tiengo Ferreira and
                  Vijay Kumar Eranti and
                  Elie Bursztein},
  editor       = {Marina Blanton and
                  William Enck and
                  Cristina Nita{-}Rotaru},
  title        = {Supporting Human Raters with the Detection of Harmful Content Using
                  Large Language Models},
  booktitle    = {{IEEE} Symposium on Security and Privacy, {SP} 2025},
  pages        = {2772--2789},
  publisher    = {{IEEE}},
  year         = {2025},
  url          = {https://doi.org/10.1109/SP61157.2025.00082},
  doi          = {10.1109/SP61157.2025.00082},
  address={San Francisco, CA, USA}
}

@article{Kumar_2024, 
  title={Watch your language: Investigating content moderation with large language models}, 
  volume={18}, 
  DOI={10.1609/icwsm.v18i1.31358}, 
  journal={Proceedings of the International AAAI Conference on Web and Social Media}, 
  author={Kumar, Deepak and AbuHashem, Yousef Anees and Durumeric, Zakir}, 
  year={2024}, 
  month={May}, 
  pages={865–878}
}

@article{11140757,
  author={Divya, P. and Samprakash, G. and Yazhini, B. and Kesavan, R. and Saravanakumar, R. and Lakshmi, S. Jeya},
  journal={2025 8th International Conference on Computing Methodologies and Communication (ICCMC)}, 
  title={AI-based Content Moderation System for Offensive Data Detection}, 
  year={2025},
  pages={1803-1809},
  doi={10.1109/ICCMC65190.2025.11140757}
}

@INPROCEEDINGS{11323612,
  author={Surya, J. Rajesh and Mai, C. Kiran and Gangappa, M.},
  booktitle={2025 IEEE 5th International Conference on ICT in Business Industry \& Government (ICTBIG)}, 
  title={Real-Time Multimodal Content Moderation and Account Suspension System}, 
  year={2025},
  pages={1-7},
  doi={10.1109/ICTBIG68706.2025.11323612},
  publisher={IEEE},
  address={Madhya Pradesh, India}
}

@misc{Corporation_2026, 
title={Roblox experiences}, 
url={https://www.roblox.com/charts?device=computer&country=us}, 
journal={Roblox}, 
author={Corporation, Roblox}, 
year={2026}
}

@misc{Singh_2025, 
  title={The infrastructure supporting record-breaking experiences}, 
  url={https://about.roblox.com/newsroom/2025/06/roblox-infrastructure-supporting-record-breaking-games}, 
  journal={Roblox}, 
  author={Anupam Singh, Senior Vice President of Engineering}, 
  year={2025}, 
  month={Jun}
}

@article{Mpofu31122026,
author = {Favourate Y. Mpofu and Samantha Lufuno Mudau},
title = {Saturation in the digital qualitative research: examining the factors influencing saturation attainment},
journal = {Cogent Arts \& Humanities},
volume = {13},
number = {1},
pages = {2639313},
year = {2026},
publisher = {Cogent OA},
doi = {10.1080/23311983.2026.2639313},
URL = {https://doi.org/10.1080/23311983.2026.2639313},
}

@misc{Chen_Anandayuvaraj_Davis_Rahaman_2024,
  url={https://arxiv.org/pdf/2310.01653},
  journal={On the contents and utility of IOT cybersecurity guidelines},
  author={Chen, Jesse and Anandayuvaraj, Dharun and Davis, James C and Rahaman, Sazzadur},
  year={2024},
  month={Jul}
}

@misc{Roblox_bert, 
  title={How We Scaled Bert to Serve 1+ Billion Daily Requests on CPUs}, 
  url={https://about.roblox.com/newsroom/2020/05/scaled-bert-serve-1-billion-daily-requests-cpus}, 
  journal={Roblox}, 
  author={Roblox Corporation}, 
  year={2020}, 
  month={May}
}

@misc{Roblox_text_filtering, 
  title={Text filtering}, 
  url={https://create.roblox.com/docs/ui/text-filtering}, 
  journal={Create.roblox.com}, 
  author={Roblox Corporation}, year={2026}
}

@misc{Roblox_chat_overview, 
  title={Text chat overview}, 
  url={https://create.roblox.com/docs/chat/in-experience-text-chat}, 
  journal={Create.roblox.com}, 
  author={Roblox Corporation}, year={2026}
}

@misc{Roblox_matchmaking, 
  title={Matchmaking}, 
  url={https://create.roblox.com/docs/matchmaking}, 
  journal={Create.roblox.com}, 
  author={Roblox Corporation}, 
  year={2026}
}

@misc{Roblox_scoring, 
  title={Server Scoring}, 
  url={https://create.roblox.com/docs/matchmaking/scoring}, 
  journal={Create.roblox.com}, 
  author={Roblox Corporation}, 
  year={2026}
}

@misc{Roblox_attributes_signals, 
  title={Attributes and Signals}, 
  url={https://create.roblox.com/docs/matchmaking/attributes-and-signals#existing-signals}, 
  journal={Create.roblox.com}, 
  author={Roblox Corporation}, 
  year={2026}
}

@misc{Roblox_player, 
  title={Player}, 
  url={https://create.roblox.com/docs/reference/engine/classes/Player#Idled}, 
  journal={Create.roblox.com}, 
  author={Roblox Corporation}, 
  year={2026}
}

@misc{Ruvalcaba_Mercer, 
  title={Adolescent Sexting, Violence, and Sexual Behaviors: An Analysis of 2014 and 2016 Pennsylvania Youth Risk Behavior Survey Data}, 
  url={https://doi.org/10.1111/josh.13290}, 
  journal={National Library of Medicine}, 
  author={Ruvalcaba, Yanet and Mercer Kollar, Laura M and Jones, Sherry Everett and Mercado, Melissa C and Leemis, Ruth W and Zhen-Qiang}, 
  year={2023}, 
  month={Aug}
}

@misc{Kabiru, 
  title={Risk and Protective Factors for the Sexual and Reproductive Health of Young Adolescents: Lessons Learnt in the Past Decade and Research Priorities Moving Forward}, 
  url={https://doi.org/10.1016/j.jadohealth.2024.03.007}, 
  journal={Journal of Adolescent Health}, 
  author={Kabiru, Caroline W and Habib, Helen H and Beckwith, Sam and Ajayi, Anthony Idowu and Mukabana, Sheila and Machoka, Beryl Nyatuga and Blum, Robert Wm and Kagesten, Anna E}, 
  year={2024}, 
  month={Oct}
}

@misc{Deathwalker_2026, 
  title={Korblox Deathwalker}, 
  url={https://www.roblox.com/bundles/319226/Korblox-Deathwalker}, 
  journal={Roblox}, 
  author={Corporation, Roblox}, 
  year={2026}
}

@misc{Wang_Rajtmajer_2025, 
  title={The unappreciated role of intent in algorithmic moderation of abusive content on social media: HKS Misinformation Review}, 
  url={https://doi.org/10.37016/mr-2020-180}, 
  journal={Misinformation Review}, 
  author={Wang, Xinyu and Koneru, Sai and Venkit, Pranav Narayanan and Frischmann, Brett and Rajtmajer, Sarah}, 
  year={2025},
  month={Jul}
}

@article{Garcia_70029,
author = {Diaz-Garcia, J. Angel and Carvalho, Joao Paulo},
title = {A Literature Review of Textual Cyber Abuse Detection Using Cutting-Edge Natural Language Processing Techniques: Language Models and Large Language Models},
journal = {WIREs Data Mining and Knowledge Discovery},
volume = {15},
number = {3},
pages = {38},
doi = {https://doi.org/10.1002/widm.70029},
url = {https://wires.onlinelibrary.wiley.com/doi/abs/10.1002/widm.70029},
eprint = {https://wires.onlinelibrary.wiley.com/doi/pdf/10.1002/widm.70029},
year = {2025}
}

@misc{Fandom_2021, 
  title={GameCharlie1}, 
  url={https://robloxcities.fandom.com/wiki/GameCharlie1}, 
  journal={Robloxiapedia}, 
  author={Fandom}, 
  year={2021}
}

@article{Mpofu_16094069251348542,
author = {Favourate Y. Mpofu},
title ={The Saturation Dilemma Reconsidered: Role, Challenges and Controversies for Qualitative Research in the Digital Era},
journal = {International Journal of Qualitative Methods},
volume = {24},
number = {},
pages = {16094069251348542},
year = {2025},
doi = {10.1177/16094069251348542},
URL = {https://doi.org/10.1177/16094069251348542},
}

@Article{educsci13020098,
AUTHOR = {Daher, Wajeeh},
TITLE = {Saturation in Qualitative Educational Technology Research},
JOURNAL = {Education Sciences},
VOLUME = {13},
YEAR = {2023},
NUMBER = {2},
ARTICLE-NUMBER = {98},
URL = {https://www.mdpi.com/2227-7102/13/2/98},
ISSN = {2227-7102},
DOI = {10.3390/educsci13020098}
}

@article{Perea_2008,
author = {Perea, Manuel and Duñabeitia, Jon Andoni and Carreiras, Manuel},
year = {2008},
month = {02},
pages = {237-241},
title = {R34D1NG W0RD5 W1TH NUMB3R5},
volume = {34},
journal = {Journal of Experimental Psychology: Human Perception and Performance},
doi = {10.1037/0096-1523.34.1.237}
}

@inproceedings{juneja2023,
author = {Juneja, Prerna and Bhuiyan, Md Momen and Mitra, Tanushree},
title = {Assessing enactment of content regulation policies: A post hoc crowd-sourced audit of election misinformation on YouTube},
year = {2023},
isbn = {9781450394215},
publisher = {Association for Computing Machinery},
address = {New York, NY, USA},
url = {https://doi.org/10.1145/3544548.3580846},
doi = {10.1145/3544548.3580846},
booktitle = {Proceedings of the 2023 CHI Conference on Human Factors in Computing Systems},
articleno = {545},
numpages = {22},
location = {Hamburg, Germany},
series = {CHI '23}
}

@article{Hong_corona,
author = {Traci Hong and Zilu Tang and Manyuan Lu and Yunwen Wang and Jiaxi Wu and Derry Wijaya},
title ={Effects of \#coronavirus content moderation on misinformation and anti-Asian hate on Instagram},
journal = {New Media \& Society},
volume = {27},
number = {2},
pages = {931-954},
year = {2025},
doi = {10.1177/14614448231187529},
URL = {https://doi.org/10.1177/14614448231187529}
}

@misc{goldstein2023,
      title={Understanding the (In)Effectiveness of Content Moderation: A Case Study of Facebook in the Context of the U.S. Capitol Riot}, 
      author={Ian Goldstein and Laura Edelson and Minh-Kha Nguyen and Oana Goga and Damon McCoy and Tobias Lauinger},
      year={2023},
      eprint={2301.02737},
      archivePrefix={arXiv},
      primaryClass={cs.SI},
      url={https://arxiv.org/abs/2301.02737}, 
}

@article{Trujillo_2025,
author = {Trujillo, Amaury and Fagni, Tiziano and Cresci, Stefano},
title = {The DSA Transparency Database: Auditing Self-reported Moderation Actions by Social Media},
year = {2025},
issue_date = {May 2025},
publisher = {Association for Computing Machinery},
address = {New York, NY, USA},
volume = {9},
number = {2},
url = {https://doi.org/10.1145/3711085},
doi = {10.1145/3711085},
journal = {Proc. ACM Hum.-Comput. Interact.},
month = may,
articleno = {CSCW187},
numpages = {28}
}
